\documentclass[useAMS,usenatbib]{mn2e}
\usepackage{txfonts}
\usepackage{natbib}
\usepackage{aas_macros}
\usepackage{graphicx}
\usepackage{subfigure}
\usepackage{mathrsfs}
\usepackage{xfrac}

\newcommand{\gsim}{\mathrel{\hbox{\rlap{\lower.55ex \hbox {$\sim$}}
                   \kern-.3em \raise.4ex \hbox{$>$}}}}
\newcommand{\lsim}{\mathrel{\hbox{\rlap{\lower.55ex \hbox {$\sim$}}
                   \kern-.3em \raise.4ex \hbox{$<$}}}}

\newcommand\density{$\rm g \: cm^{-3}$}
\newcommand\sdensity{$\rm g \: cm^{-2}$}
\newcommand\mathdensity{{~\rm g \: cm^{-3}}}

\newcommand\rhill{$R_{\rm H}$}

\newcommand\rorbit{$r_{\rm p}$}

\newcommand\mrhill{ R_{\rm H} }

\newcommand\solarmass{$\rm M_{\odot}$}

\newcommand\earthmass{$\rm M_{\oplus}$}
\newcommand\jupitermass{$\rm M_{Jupiter}$}

\title[Hydrodynamic collapse models]{The growth and hydrodynamic collapse of a protoplanet envelope}
\author[B.A. Ayliffe \& M.R. Bate] {Ben A. Ayliffe$^{1,2}$\thanks{E-mail:ayliffe@astro.ex.ac.uk}
and Matthew R. Bate$^{1}$\\
$^1$School of Physics, University of Exeter, Stocker Road, Exeter EX4 4QL\\
$^2$Monash Centre for Astrophysics (MoCA) \& School of Mathematical Sciences, Monash University, Clayton, Vic 3800, Australia}

\date{\today}
\begin{document}
\maketitle

\begin{abstract}
We have conducted three-dimensional self-gravitating radiation hydrodynamical models of gas accretion onto high mass cores (15-33~\earthmass) over hundreds of orbits. Of these models, one case accretes more than a third of a Jupiter mass of gas, before eventually undergoing a hydrodynamic collapse. This collapse causes the density near the core to increase by more than an order of magnitude, and the outer envelope to evolve into a circumplanetary disc. A small reduction in the mass within the Hill radius (\rhill) accompanies this collapse as a shock propagates outwards. This collapse leads to a new hydrostatic equilibrium for the protoplanetary envelope, at which point 97 per cent of the mass contained within the Hill radius is within the inner 0.03~\rhill \ which had previously contained less than 40 per cent. Following this collapse the protoplanet resumes accretion at its prior rate. The net flow of mass towards this dense protoplanet is predominantly from high latitudes, whilst at the outer edge of the circumplanetary disc there is net outflow of gas along the midplane. We also find a turnover of gas deep within the bound envelope that may be caused by the establishment of convection cells.
\end{abstract}


\section{Introduction}

The ideas discussed in this paper begin with \cite{PerCam1974}, who stated that ``when the mass of the core becomes sufficiently great, the surrounding gaseous envelope will become hydrodynamically unstable against collapse onto the planetary core''. This process is controlled by the battle between gravity that acts to contract an envelope onto the core and the gas pressure which acts to support it.

\cite{Miz1980} performed stability calculations to determine the combinations of core masses and opacities that would make a protoplanetary envelope unstable to collapse. Work of his contemporaries considering the structure of giant planets in the solar system suggested that each such planet (Jupiter, Saturn, Uranus, \& Neptune) possessed a solid core with a mass of order 10~\earthmass \ \citep{Sla1977,HubMac1980}. Using these values as a target, and assuming a fixed accretion rate, \citeauthor{Miz1980} concluded that a grain opacity of $\kappa \approx 1~{\rm cm^2 \: g^{-1}}$ was required in the envelope material during formation to trigger a collapse when the core mass was $\approx 10$~\earthmass. Lower opacities were found to lead to envelope collapse at lower core masses. Following the envelope collapse \citeauthor{Miz1980} states that continued accretion is likely required for the protoplanet's to attain their final masses.

It was suggested later by \cite{BodPol1986}, who performed evolution calculations that included evolution beyond the attainment of the critical core mass, that rather than a dynamical collapse, the envelope may quasi-statically contract onto the solid core. They suggested that if a sufficient mass of molecular hydrogen in the envelope was apt to undergo dissociation, then this would remove enough energy from the contraction to bring about a dynamical collapse. However, their models indicate that such dissociative regions possess an insufficient fraction of the envelope mass for this to occur.

In the early 1990s \citeauthor{Wuc1990} wrote a series of papers exploring the evolution of a protoplanetary envelope through the hydrostatic phases approaching the critical core mass, and in what he found to be a subsequent hydrodynamic phase \citep{Wuc1991}. Solving the equations of radiation hydrodynamics in one-dimension, \citeauthor{Wuc1991} indicated that following a period of quasi-static contraction, during which the envelope heats up, the transport of this heat out through the envelope by convection and radiation perturbs the hydrostatic equilibrium. In particular \citeauthor{Wuc1991} cites the $\kappa-\rm{mechanism}$ as the means of exciting the dynamical waves that destabilise the envelope. The result of this hydrodynamic evolution was the ejection of a large fraction of the envelope mass, rather than an inwards collapse as had previously been supposed by \cite{PerCam1974}.

\cite{PolHubBodLis1996} continued performing models assuming a quasi-hydrostatic contraction, and suggested that further work was required to consider the hydrodynamic evolution of young protoplanets to establish the proper evolution scenario. A step in this direction was taken by \cite{TajNak1997} who performed a stability analysis of a growing protoplanetary envelope using a distinct numerical code to that of previous authors. Their aim was to determine whether quasi-static contraction, or an envelope instability akin to that suggested by \citeauthor{Wuc1991}, was the more likely evolutionary course for a growing protoplanet. They perturbed the envelope at intervals during its evolution to see if such action might push a marginally stable system towards instability. They concluded that quasi-static contraction was viable for a protoplanet growing all the way to a Jupiter mass.

There has since been a substantial amount of work considering the gas accretion rates that cores might achieve in circumstellar discs with a variety of properties. \cite{IkoNakEmo2000} performed quasi-static evolutionary models to determine the dependence of gas accretion upon core mass, grain opacity, and the core's accretion history, finding that these factors were strongly, moderately, and weakly significant respectively. \cite{BryCheLinNel1999} and \cite{LubSeiArt1999} performed locally-isothermal hydrodynamics simulations of discs containing planetary cores, the latter finding that the accretion rate drops off as the protoplanet mass becomes very large; a result of the broadening disc gap that it forms. Further hydrodynamic models with more realistic thermodynamics were performed by \cite{DAnHenKle2003}, \cite{KlaKle2006}, \cite{PaaMel2008}, and \cite{AylBat2009}, finding similar turn overs in the accretion rate with increasing mass, and illustrating the impact of grain opacity on accretion. These models have generally had limited resolution in the vicinity of the protoplanet, and though \cite{AylBat2009} achieved high resolution, the evolutionary period was extremely short due to the computational demands of the calculations.

In this paper we report results from three-dimensional self-gravitating radiation hydrodynamics calculations that resolve the protoplanet's envelope, whilst modelling its hydrodynamic evolution and growth within a section of a circumstellar disc. Using high mass discs (though still stable; Toomre Q $>> 1$), and assuming low opacities, we achieve accretion rates that allow significant envelope growth in only a few hundred orbits, allowing us to examine their development. Our computational method is described in Section~\ref{sec:setup}, followed by our results in Section~\ref{sec:results}, and a discussion of their relationship with previous results in Section~\ref{sec:discussion}.

\section{Computational Method}
\label{sec:setup}

The calculations discussed in this paper were performed using a three-dimensional SPH code. This SPH code is derived from a code first developed by \citeauthor{Ben1990} (\citeyear{Ben1990}; \citealt{BenCamPreBow1990}) which has undergone substantial modification in subsequent years. Energy and entropy are conserved to timestepping accuracy by use of the variable smoothing length formalism of \cite{SprHer2002} and \cite{Mon2002}, where our particular implementation is described in \cite{PriMon2007}. Gravitational forces are calculated and particle neighbours are found using a binary tree. Radiative transfer is modelled using the flux-limited diffusion approximation, employing the method developed by \citet*{WhiBatMon2005} and \citet{WhiBat2006}. Integration of the SPH equations is achieved using a second-order Runge-Kutta-Fehlberg integrator with particles having individual timesteps \citep{Bat1995}. Gas within the models is subject to an artificial viscosity, implemented in a parameterised form as developed for SPH by \cite{MonGin1983}, and modified to deal with high Mach number shocks by \cite{Mon1992}. The code has been parallelised by M. Bate using OpenMP and MPI.

\subsection{Model setup}
\label{sec:model}

The calculations we have performed were conducted by modelling a small section of a circumstellar disc, centred upon a protoplanetary core and corotating with its orbit. The protoplanet is modelled as a gravitating mass with a `surface'.  It attracts gas from the disc, building up an atmosphere which is supported by the surface.
The disc section measures $r=1\pm 0.15~r_{\rm p}$ ($5.2 \pm 0.78~{\rm AU}$), and $\phi = \pm~0.15$ radians. The protoplanet is sited at a radius of \rorbit, which in all cases is equivalent to 5.2 AU, and orbits about a star of 1~\solarmass. The disc has a surface density profile of $\Sigma \propto r^{-1/2}$, and a temperature profile of  $T_{\rm g} \propto r^{-1}$ (giving a scaleheight of $H/r=0.05$), where these profiles are equivalent to those used in our previous work, and were originally chosen to match work of \citet{LubSeiArt1999} and \citet{BatLubOgiMil2003}. The initial temperature at \rorbit \ is $\approx 73$~K, where this temperature is taken to principally result from stellar irradiation. Near the midplane of the disc, this initial temperature may increase due to viscous heating as the model evolves if the opacity is sufficient to insulate the region against rapid radiative cooling. However, at the upper and lower radiation boundaries (see section~\ref{method:thermo} below), where the medium is always optically thin, an assumption of temperatures dominated by the stellar irradiaton is reasonable. The surface density at \rorbit \ has an unperturbed value of 750~\sdensity, which gives a relatively massive disc, comprising 0.1~\solarmass \ of gas within 25~AU.  { We use a high disc surface density because \cite{AylBat2009} found that increasing the disc surface density resulted in somewhat faster accretion rates onto embedded protoplanets and the goal of this paper is to investigate the three-dimensional evolution of protoplanets that accrete massive gaseous envelopes.}

Further details of our model setup are given below, but the method is identical to that employed in \cite{AylBat2009}, and that paper contains somewhat more extensive information, including resolution tests. We note that the most interesting model discussed in this paper, a case which results in a hydrodynamic envelope collapse, required more than 6 CPU years to reach its final state. It is this extensive calculation time which has limited the number of models that is has been possible to perform.

\subsection{Disc sections}

Within the disc section being modelled, the particles are initially distributed according to the underlying density profile, and their velocities are Keplerian. The protoplanet is orbiting its star in an anticlockwise direction, and in the corotating frame of the modelled section this leads particles at $r < r_{\rm p}$ to orbit anticlockwise, and those at $r > r_{\rm p}$ to orbit in a clockwise fashion. Particles encountering the boundary of the section are removed from the calculation, whilst a number of ghost particles beyond the domain of the calculation, along its boundaries, act to replicate the pressure and viscous forces expected from a continuous disc. To prevent the depletion of gas within the disc section, particles are injected along the boundaries where the material should flow in. For the anticlockwise orbiting gas this input is along the $\phi = -0.15$~radian boundary, between $r = 0.85 - 1.0$~\rorbit, whilst the clockwise flowing gas is injected along the $\phi = 0.15$~radian boundary, between $r = 1.0 - 1.15$~\rorbit. The injection scheme does more than simply replace gas which leaves the section. The velocity and density structure of the injected gas is obtained from three-dimensional global simulations of protoplanets embedded in discs performed using ZEUS \citep{BatLubOgiMil2003}, such that a gap is opened in the disc which corresponds to the mass of the embedded protoplanet. The models of \citeauthor{BatLubOgiMil2003} used a surface density of 75~\sdensity, but the particles injected in the calculations discussed here have their masses scaled to reflect the chosen surface density of 750~\sdensity. As the protoplanet grows the gap width is suitably increased by interpolation through the protoplanet mass range provided by \cite{BatLubOgiMil2003} (1~\earthmass  $-1~ {\rm M_{Jupiter}}$). The injected particles come to dominate the section after less than five orbits, ensuring the structure is consistent with the presence of the protoplanet.

\subsection{Thermodynamics}
\label{method:thermo}

All the calculations discussed in this paper were performed using radiation hydrodynamics, employing a two temperature (gas and radiation) radiative transfer scheme using a flux-limited diffusion approximation (as described by \citealt{WhiBatMon2005} and \citealt{WhiBat2006}). Work and artificial viscosity act to increase the thermal energy of the gas, and work done on the radiation field increases the radiative energy, which can be transported via flux-limited diffusion. The energy transfer between the gas and radiation fields is dependent upon their relative temperatures, the gas density, and the gas opacity.

The gas is treated using an ideal gas equation of state $p=\rho T_{g} R_{g}/\mu$ where $R_{g}$ is the gas constant, $\rho$ is the density, $T_{g}$ is the gas temperature, and $\mu$ is the mean molecular mass. The thermal evolution takes into account the translational, rotational, and vibrational degrees of freedom of molecular hydrogen (assuming a 3:1 mix of ortho- and para-hydrogen; see \citealt{BolHarDurMic2007}). Also included are molecular hydrogen dissociation, and the ionisations of hydrogen and helium. The hydrogen and helium mass fractions are $X = 0.70$ and $Y = 0.28$, respectively, whilst the contribution of metals to the equation of state and the thermal evolution is neglected.

The flux-limited diffusion scheme transfers energy between SPH particles, which does not enable it to radiate into a vacuum. In order that the disc can cool from its upper and lower surfaces, a boundary is applied that maintains the initial temperature profile in the high atmosphere of the disc. This boundary is situated at a height above/below the midplane that corresponds to the edge of the optically thick region, that is where the optical depth ($\tau$) from outside the disc to that depth is $\tau \approx 1$. SPH particles comprising the boundary regions evolve normally, but their energies are set according to the initial radial profile, allowing them to act as energy sinks.

\subsection{Opacity treatment}

We use the opacity tables of \cite{PolMcKChr1985} to provide grain opacities, whilst the tables of \cite{Ale1975} (the IVa King model) provide the gas opacities at higher temperatures when the grains have sublimated. The former table gives interstellar grain opacities (IGO) for solar metallicity molecular gas, but in this work we reduce these opacities by orders of magnitude below this nominal level; we divide by factors of 100 and 1000. The justification for this comes of the likely agglomeration or sublimation of grains in the vicinity of a forming protoplanet \citep{Pod2003,MovBodPodLis2010}. We do not modify the gas opacities, but enforce a minimum for the grain opacities at the interface between the two regimes that corresponds to the gas minimum ensuring a smooth transition (see \citealt{AylBat2009} for more details).

\subsection{Planetary Cores}
\label{section:cores}

The planetary cores in these simulations are modelled by a gravitational potential, and a surface potential that yields an opposing force upon gas within one core radius of the core's surface. The combination of the gravitational and surface forces takes the form of a modification to the usual gravitational force as
\begin{equation}
F_{r} = - \frac{{\rm G}M_{\rm c}}{r^{2}}\left(1 - \left(\frac{2 R_{\rm c}-r}{R_{\rm c}}\right)^{4}\right)
\label{eq:surface}
\end{equation}
for $r <2~R_{\rm c}$ where $r$ is the radius from the centre of the planetary core, $R_{\rm c}$ is the radius of the core, and $M_{\rm c}$ is the mass of the core (see  \citealt{AylBat2009} for further details).  This equation yields zero net force between a particle and the planetary core at the surface radius $R_{\rm c}$, whilst inside of the core's radius the force is outwards and increases rapidly with decreasing radius. Gas particles therefore come to rest very close to the core radius, though the equilibrium position is slightly inward of this value due to the pressure exerted by the gas that accumulates on top of the inner most layer of particles.

\cite{SeaKucHieMil2007} calculate core radii for solid exoplanets, amongst which are cases comprising of 75 per cent water, 22 per cent silicates, and 3 per cent iron. We use these models to determine the realistic sizes of protoplanetary cores that correspond to the masses used in this paper. These core radii were employed for a number of calculations in \cite{AylBat2009}, but these cases were not evolved for many orbits due to the short timesteps required deep within the planetary potential. To follow the evolution over longer periods it has been necessary to perform calculations using larger core radii. To this end we scale up these previously used core radii by factors of 3 or 10 to reduce the required computation time. We also performed new models using the realistic core radii, but these calculations were too slow to give useful results and thus are not reported here. The properties of our different models are given in Table~\ref{table:models}. { The rate at which the planetary core accretes gas may be affected by the form of the surface potential described by equation~\ref{eq:surface}, and this is explored briefly in Section~\ref{sec:accretion}.} A smooth start to the calculations is provided by shrinking exponentially towards the desired $R_{\rm c}$ from an initial radius of 0.01~$r_{\rm p}$ over the course of the first orbit.

\renewcommand{\tabcolsep}{1.5mm}
\begin{table}
\centering
\begin{tabular}{c l c c c c c}
\hline\hline
Model & \multicolumn{2}{c}{Core mass}  & Core radius &  $\times$ Physical & Opacity \\ 
& $[{\rm M}_{\odot}] $ & $[{\rm M}_{\earth}]$ & [${\rm r}_{\rm p}$] & core size & [\% IGO] \\ 
\hline
A & $4.54 \times 10^{-5} $  & 15 & $2.2 \times 10^{-4}$ & 10 & 0.1\\
B & $6 \times 10^{-5} $  & 20 & $6.6 \times 10^{-5}$ &  3 & 0.1\\
C & $6 \times 10^{-5}$  & 20 & $6.6 \times 10^{-5}$ &  3 & 1\\
D & $6 \times 10^{-5}$  & 20 & $2.2 \times 10^{-4}$ &  10 & 0.1 \\
E &$6 \times 10^{-5}$  & 20 & $2.2 \times 10^{-4}$ &  10 & 1 \\
J & $1 \times 10^{-4}$  & 33 & $2.54 \times 10^{-4}$ & 10 & 0.1 \\[1ex]
\hline
\end{tabular}
\caption{Properties of the various models described in this paper, all of which were performed in a disc with a surface density of 750~\sdensity \ at \rorbit. The opacity is given as a percentage of the interstellar grain opacity (IGO) assumed in the opacity tables we employ. Core radii are all based on the models of \citealt{SeaKucHieMil2007}, multiplied by factors of 3 and 10 as marked.}
\label{table:models}
\end{table}

\begin{figure}
\centering
\includegraphics[width=\columnwidth]{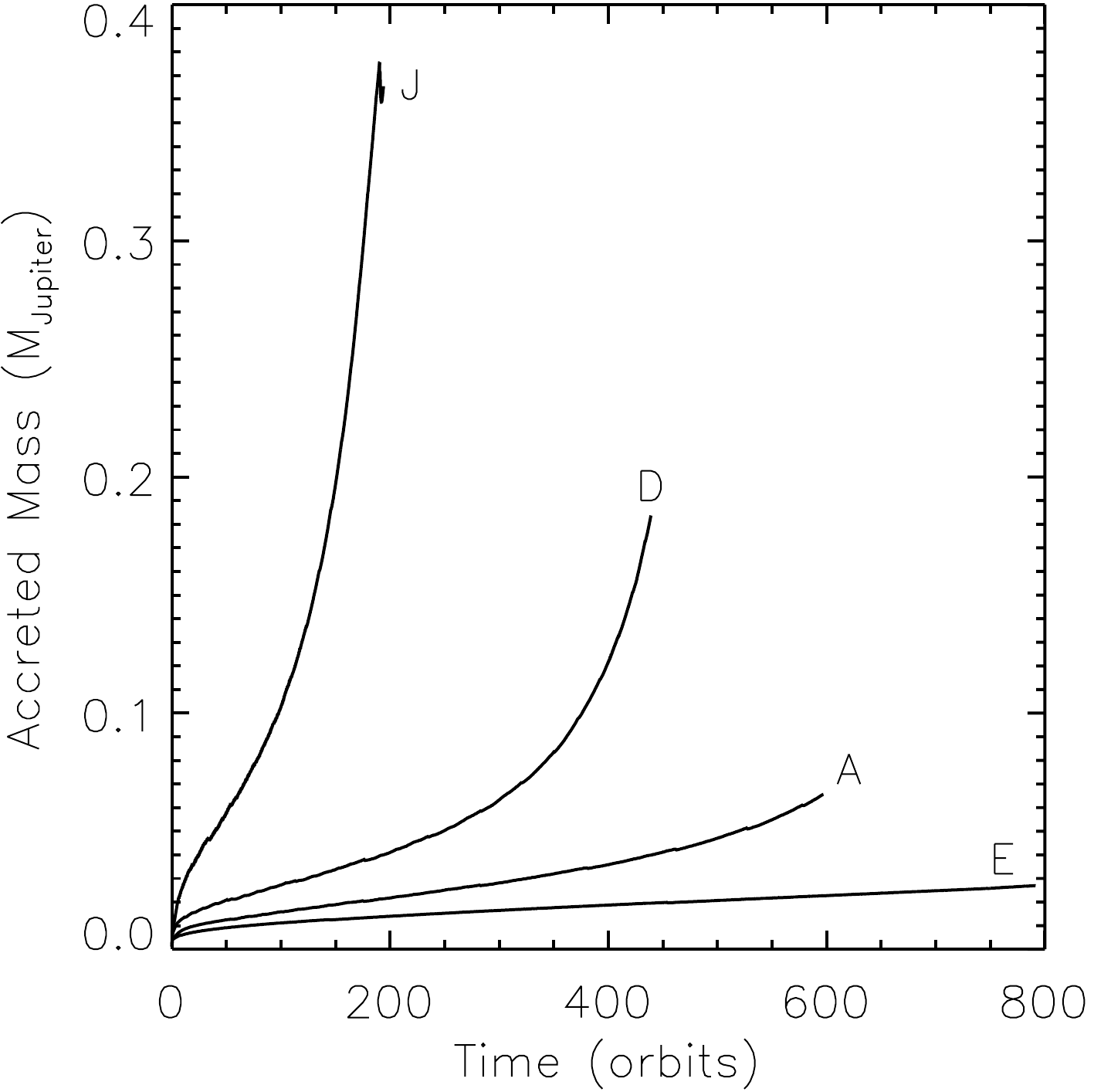}
\caption{Accretion histories for protoplanets with initial core masses of 15 (A), 20 (D, E), and 33 ( J)~\earthmass.  The models possess core radii 10 times their likely physical size. Model E is performed using a disc with 1 per cent interstellar grain opacity, whilst all the others use 0.1 per cent; model E is otherwise identical to D. The size and opacity dependencies are further illustrated in Fig.~\ref{fig:m20accretion}.}
\label{fig:10Raccretion}
\end{figure}

\subsection{Measuring the gas accretion rates onto the planetary cores}

We measure the gas accretion rates by calculating the rate at which gas passes into the self-consistently calculated Hill sphere of the protoplanet given by
\begin{equation}
R_{\rm H}=\sqrt[3]{\frac{M_{\rm p}}{3M_{*}}}r_{\rm p}
\label{eq:hill}
\end{equation}
where $M_{\rm p}$ is the protoplanet mass which is the sum of the core mass ($M_{\rm c}$) and the accreted mass ($M_{\rm acc}$), where accreted mass comprises all the gas within $R_{\rm H}$. The gas mass is discretised amongst the SPH particles, allowing iteration through equation~\ref{eq:hill} until such time as the addition of a particle's mass to $M_{\rm acc}$ no longer increases $R_{\rm H}$ sufficiently to encompass the next available particle.

The net flux through the Hill radius corresponds to the growth rate of the envelope and/or circumplanetary disc, which are the only repositories for gas that fails to remerge from this region. Our use of the Hill radius to measure the mass growth is arbitrary, but reasonable, since any protoplanet must be smaller than the Hill radius. It is important to note that the Hill radius does not define the extent of the envelope, which tends to be smaller (e.g. $\sim 0.25~R_{\rm H}$ \citealt{LisHubDAnBod2009}). However, the difference between the mass accretion rates (and total accreted masses) as measured at the Hill radius and at $0.3$~R$_{\rm H}$ is small except at the start of the calculation (Section \ref{sec:collapse}).

\begin{figure}
\centering
\includegraphics[width=\columnwidth]{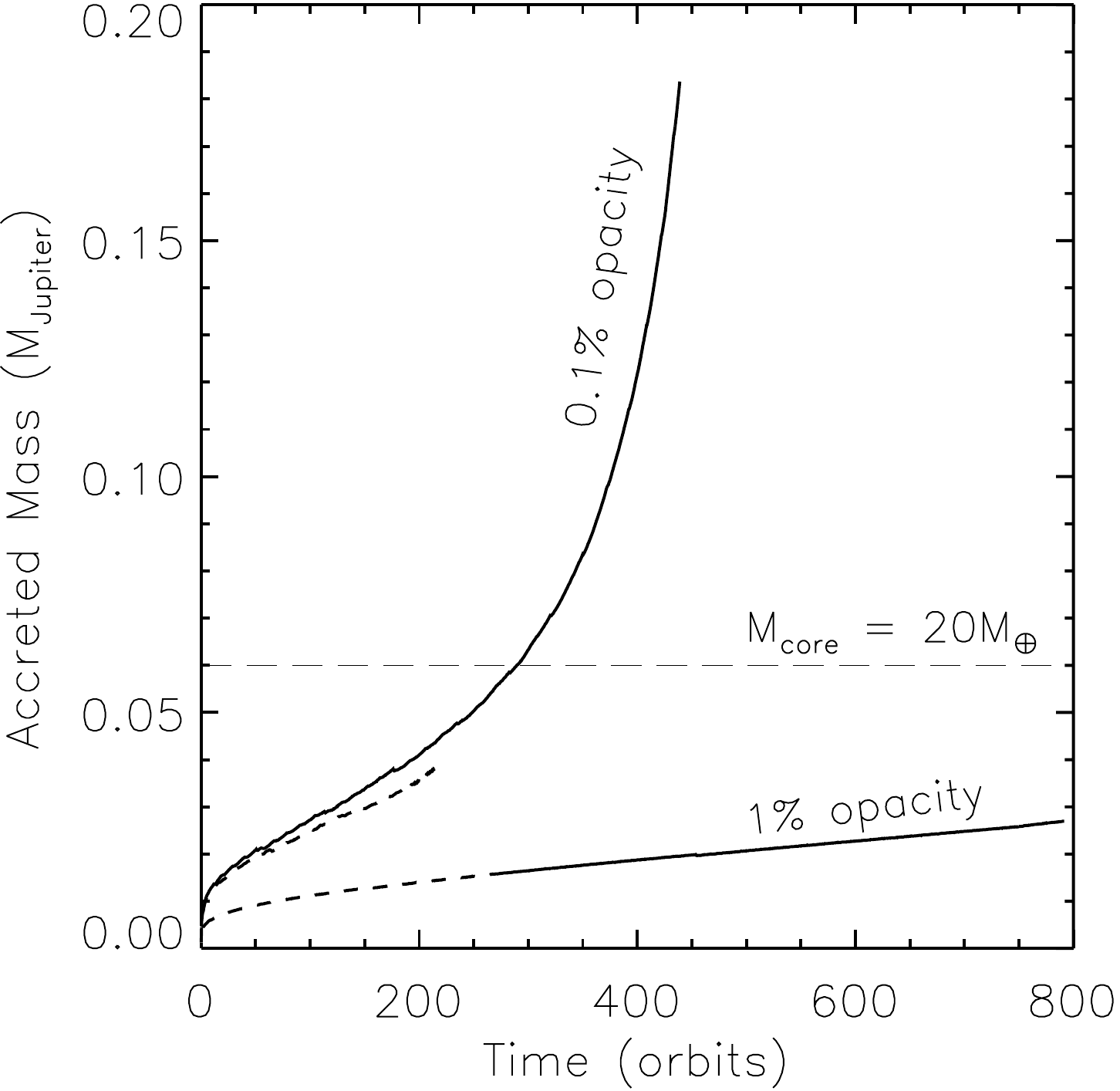}
\caption{Accretion history for a 20~\earthmass \ core with 0.1 per cent and 1 per cent opacities (as marked), for two different core radii, $3 \times$ physical (dashed lines, models B \& C), and $10 \times$ physical (solid lines, models D \& E). Note that for the higher opacity case, which accretes more slowly, the models of different assumed core radii coevolve. { For the lower opacity cases, which accrete large envelopes much more rapidly, the case with the smaller core radius accretes more slowly that the large core case during the initial $\sim 50$ orbits. However, thereafter, when their accretion rates are measured at equivalent masses, they exhibit very similar rates.}}
\label{fig:m20accretion}
\end{figure}

\section{Results}
\label{sec:results}

We have performed three-dimensional radiation hydrodynamics models of the accretion of gas onto planetary cores (or embryos) with a range of masses, over hundreds of orbits, in discs of varying opacity. This work extends upon models we performed in \cite{AylBat2009} where we followed the accretion for a relatively short period. Moreover, by using discs with highly reduced opacities, the models presented here include the accretion of much more significant envelope masses. In one case the accreted mass is sufficient to trigger a hydrodynamic collapse of the envelope (Section~\ref{sec:collapse}), followed by a return to steady gas accretion.

\subsection{Envelope accretion}
\label{sec:accretion}

{ 
We performed calculations starting with various core masses ranging from 15 $-$ 33~\earthmass \ (Table \ref{table:models}). The principal property that controls the accretion rate of a protoplanet of a given mass is the opacity of its envelope. A lower opacity enables more rapid radiative cooling, allowing the envelope to contract more quickly and so accrete gas at a faster rate \citep[]{HubBodLis2005,PapNel2005,AylBat2009}. The models presented in this work exploit this dependence to accelerate the growth process by adopting reduced opacities. The impact of opacity on the growth rate of a protoplanet is demonstrated by comparing Models E and D in Figs.~\ref{fig:10Raccretion} and \ref{fig:m20accretion}. Model E employs an opacity ten times larger than D, which results in a much slower rate of accretion for the former.
}

\begin{figure}
\centering
\includegraphics[width=\columnwidth]{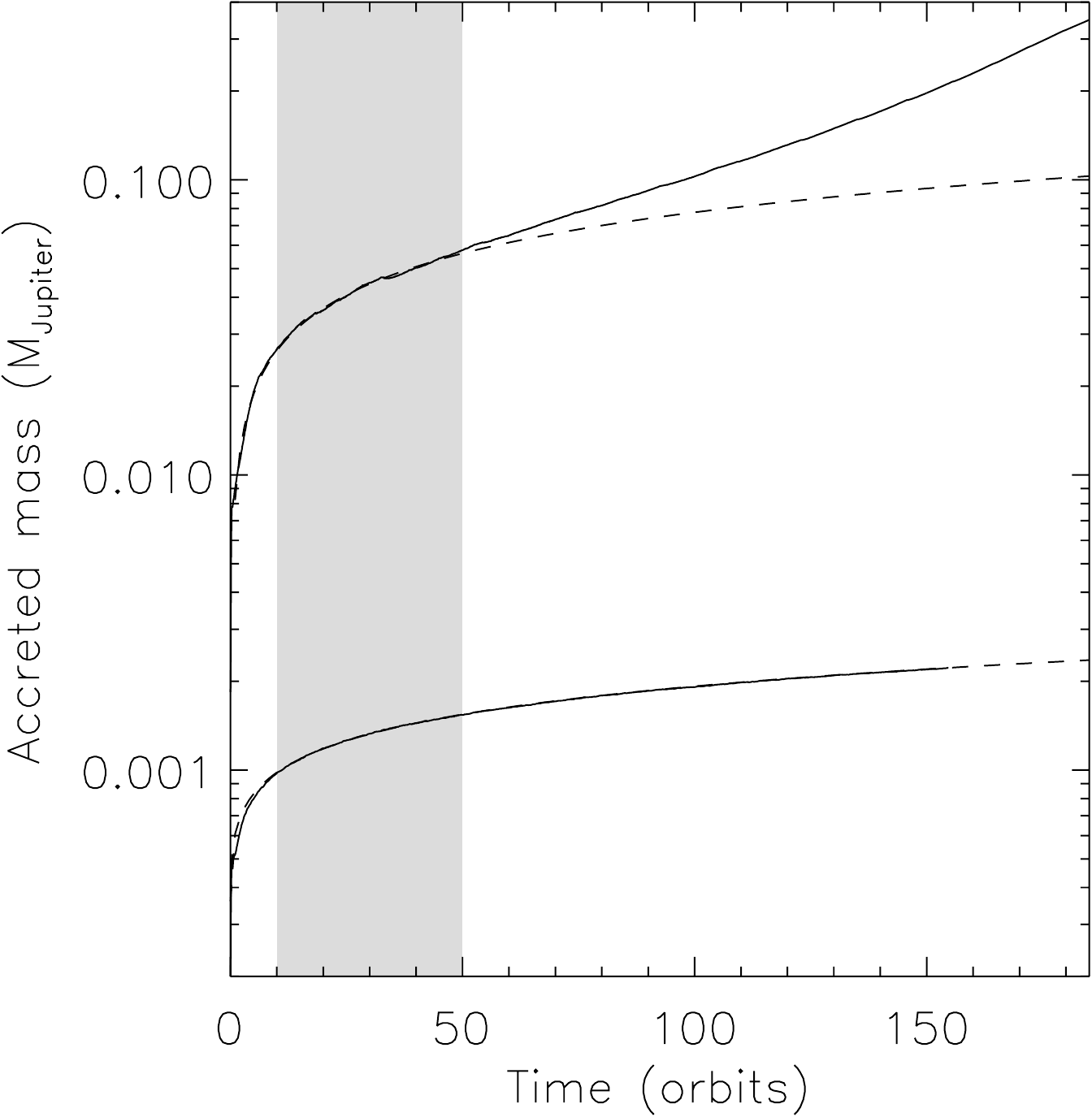}
\caption{A comparison between results from the lower surface density, shorter evolution calculations presented in \protect \cite{AylBat2009} and the longer, more rapidly accreting models discussed in this paper. The lower solid line is the accretion history over 160 orbits for a 33~\earthmass \ ($0.1~{\rm M_{Jupiter}}$) protoplanet in a disc with $\Sigma_{\rm p} = 75$~\sdensity, and interstellar grain opacity. Overlaid is a fit of the form $M = at^b + c$ (dashed line), where the fit is made over the range $10-50$ orbits (shaded) and matches the measured growth over the entire evolutionary period. The upper line is the accretion history for a 33~\earthmass \ protoplanet, in a disc of $\Sigma_{\rm p} = 750$~\sdensity, with 0.1 per cent IGO (model J), which by virtue of these conditions accretes much more rapidly. A fit to this curve is also shown. The departure of the evolutionary model from the fitted curve is as discussed in \protect \cite{AylBat2009}. The accretion rate initially declines as gas accretion builds a more optically thick envelope, and trapped heat prevents the envelope contracting, slowing its further accretion. However, when a sufficient mass is accreted, gravity comes to dominate the growth process, resulting in an accelerating accretion rate.}
\label{fig:extrap}
\end{figure}

{ 
In each case we introduced a bare core with no preexisting envelope. This results in a rapid initial gas accretion rate to form a quasi-static envelope, followed by a period of slowing accretion.  This initial growth phase is not realistic for a protoplanet that forms in situ and concurrently accretes both solids and gas during the core formation phase \citep[e.g.][]{PolHubBodLis1996, AliMorBenWin2005}.  In models that account for both the core and envelope growth, the gas accretion rate tends to have an extended period during which it is almost constant.  However, such an initially declining accretion rate was also seen in the semi-analytic models of \cite{PapNel2005}, where the planetary accretion rates initial decreased due to the significant liberation of binding energy as mass fell into the planet's potential. This energy release heats the forming envelope, increasing the pressure, and reducing the rate at which mass can be added. Eventually, however, in all models, once a critical mass is reached, the energy release within the envelope is insufficient to further retard the accretion, which then accelerates; this marks the transition from the thermally-dominated to gravitationally-dominated accretion regime. 

In \cite{AylBat2009}, we studied the initially declining accretion rate in one case which we followed for 160 orbits, but which did not reach the gravitationally-dominated regime.  We found that the slowing envelope growth could be represented very well with an analytic fit of the form $M=at^{b}+c$ where $b=0.40$.  In our new models, Model J is similar to the model from \cite{AylBat2009}, but with a much lower opacity and a higher disc surface density. This produces higher accretion rates and, thus, allows us to reach the gravitationally dominated growth regime. We refit the old model using only orbits $10-50$, and obtain essentially the same fit we obtained in \cite{AylBat2009} using all 160 orbits; this is evident in Fig.~\ref{fig:extrap} which illustrates that the fit is still extremely good at 160 orbits. Fitting Model J over the same 40 orbits, where an exponent $b = 0.45$ is found to be best, the trend of an initially reducing rate of accretion falters at the limit of the fitting region, around 50 orbits. At this point the accreted mass is a little more than half the core mass, and gravity begins to dominate the growth, leading to accelerating accretion.
}

In all cases, such as the models shown in Fig.~\ref{fig:10Raccretion}, the transition from a slowing rate of accretion to an accelerating rate of accretion occurs once the envelope mass has reached half the core mass. This begins the process of runaway accretion, that continues until such time as the envelope potentially undergoes a dynamic collapse, as will be discussed in the following section. It is this period of accretion that leads to the most significant growth of the protoplanet's mass.

{ 
As mentioned in Section \ref{section:cores}, in order to make these three-dimensional models computationally viable, we were forced to adopt non-realistic core radii for the planetary surfaces. In \cite{AylBat2009} we compared the accretion rates achieved in models using different core radii and found that for high opacities the accretion rates did not depend significantly on the core radius that was used, but for low opacities (1 per cent and 0.1 per cent interstellar grain opacities) the accretion rates obtained with smaller cores were significantly lower than for equivalent models adopting larger cores. However, the earlier models were only evolved for 10 orbits, which is still during the initial phase of envelope creation when the accretion rate is decreasing. Fig.~\ref{fig:m20accretion} illustrates the growth of 20~\earthmass \ cores embedded in a disc of either 0.1 or 1 per cent interstellar grain opacity (as marked). The solid lines present models with planetary cores 10 times the realistic core radii, and the dashed lines 3 times. Over the course of the initial settling period the different core radii in the 0.1 per cent opacity models lead to some divergence in the growth, as is evident. However, measuring the accretion rates of the two calculations at equivalent masses beyond the initial 50 orbits, it is found that these rates deviate by no more than 20 per cent from one another.  In the 1 per cent opacity case, the results obtained using the two different core radii are indistinguishable from each other. { We believe that measuring the accretion rates at longer times, when the models are more established is responsible for the relatively small differences seen with different core radii. However in this case we are only varying the core radius by just over a factor of 3, whilst in \cite{AylBat2009} the comparison was made spanning more than a factor of 12. At the present time we do not have any models with which we can ascertain to what extent the accretion rate differences measured in this previous work were due to the large factor difference in the core radius, and what fraction came of the early times at which the measurements were made. As such, the enlarged radii of the cores that we have adopted here should be kept in mind.}

{ 
In the rest of this paper we focus on one particular case, Model J, that accretes a very significant envelope in a short period of time that undergoes a dynamic collapse.  The other calculations discussed in this section are ongoing, and will eventually allow us to explore the evolution of envelopes that are built up under less extreme conditions (i.e. with slower accretion rates).
}

\begin{figure}
\centering
\includegraphics[width=\columnwidth]{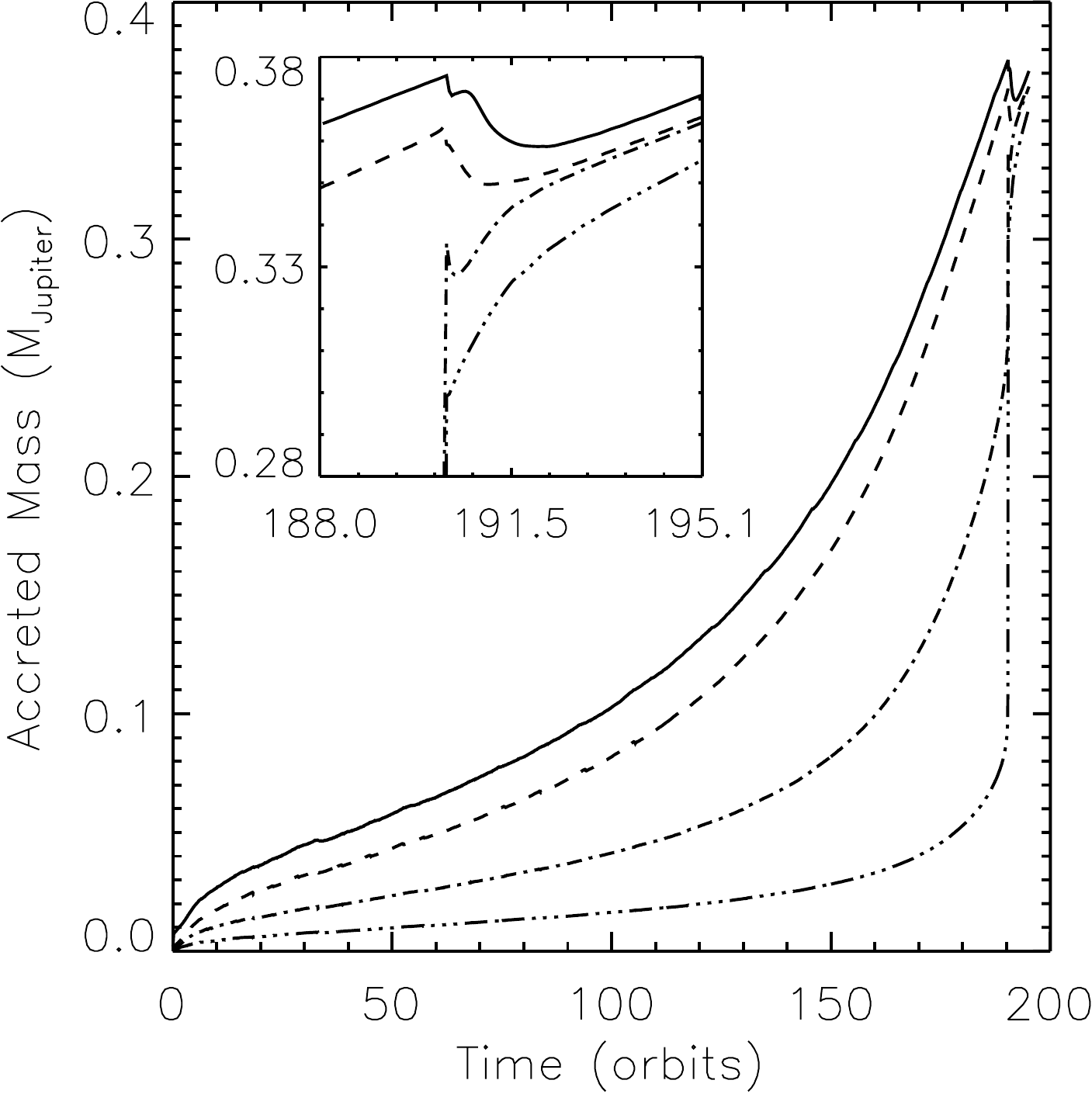}
\caption{The accretion history of the gaseous envelope in model J, where the accreted mass is that contained within the self-consistently calculated Hill radius (solid line). Also shown is the mass evolution with time within radii of 0.3 (dashed), 0.1 (dot-dash), and 0.03~\rhill \ (dots-dash). The envelope undergoes a hydrodynamical collapse after around 190 orbits when its mass is $\approx 0.375$~\jupitermass, with the resulting shock pushing material out of the Hill radius, whilst the remaining mass becomes more centrally condensed. This phase is shown in the inset panel, which illustrates the central condensation by the increasing mass within small radii as the overall mass falls. Following the collapse, accretion resumes, replacing the mass lost due to the shock propagation.}
\label{fig:racc}
\end{figure}

\subsection{Hydrodynamical collapse - Model J}
\label{sec:collapse}

Model J accretes the most significant mass of any of our models, as is to be expected given its favourable conditions. The 33~\earthmass \ core is the most massive that we employ, and in this case is coupled with a large 10 times realistic core radius, and a 0.1 per cent interstellar grain opacity. The mass evolution of this protoplanet is shown in Fig.~\ref{fig:racc}. As in all cases, the accretion rate initially drops away over the first $\sim 50$ orbits as the model settles, before beginning to increase. The acceleration of accretion becomes more marked at the crossover mass of $\approx 0.1~{\rm M_{Jupiter}}$, at around 100 orbits and continues during the runaway growth phase. A change occurs at around 190 orbits, when the accreted mass has reached $\approx 0.375~{\rm M_{Jupiter}}$. At this point the mass of the envelope is such that it overcomes the pressure gradient which hitherto had supported it, and only by centrally condensing can a new hydrostatic equilibrium be established. To achieve this the envelope rapidly collapses, in the process triggering a shock wave which propagates outwards from the very dense core that forms. This shock pushes material outwards, removing a small fraction of the material from within the Hill radius very rapidly at $\approx 190.5$~orbits. The drop in the mass within 1, 0.3, and 0.1~\rhill \ can be seen clearly in the inset panel of Fig.~\ref{fig:racc}. Most interestingly, the reduction of the mass within 0.1~\rhill \ indicates that there is a small loss of bound material as this radius falls well within the expected limit of a protoplanetary envelope. The magnitude of the mass loss from each volume is $\sim 1$~per cent of the mass therein. Following a brief increase ($\approx 190.6 - 191$~orbits), the mass within \rhill \ continues to decline over the course of a further orbit before it begins to grow once again. However, at the same time the mass within 0.1~\rhill \ increases rapidly, showing that the available material is quickly condensing around the core. Beyond $\approx 192$~orbits the accretion rate measured at the Hill radius resumes its pre-collapse rate of $4 \times 10^{-4} ~{\rm M_{Jupiter}~year^{-1}}$. Moreover, this rate appears to be consistent at nearly all radii, indicating that gas is falling directly onto the new denser envelope, rather than becoming suspended in a more extended structure as was the case prior to collapse where the accretion rates were different at different radii.

\begin{figure*}
\centering
\subfigure 
{
    \includegraphics[width=\columnwidth]{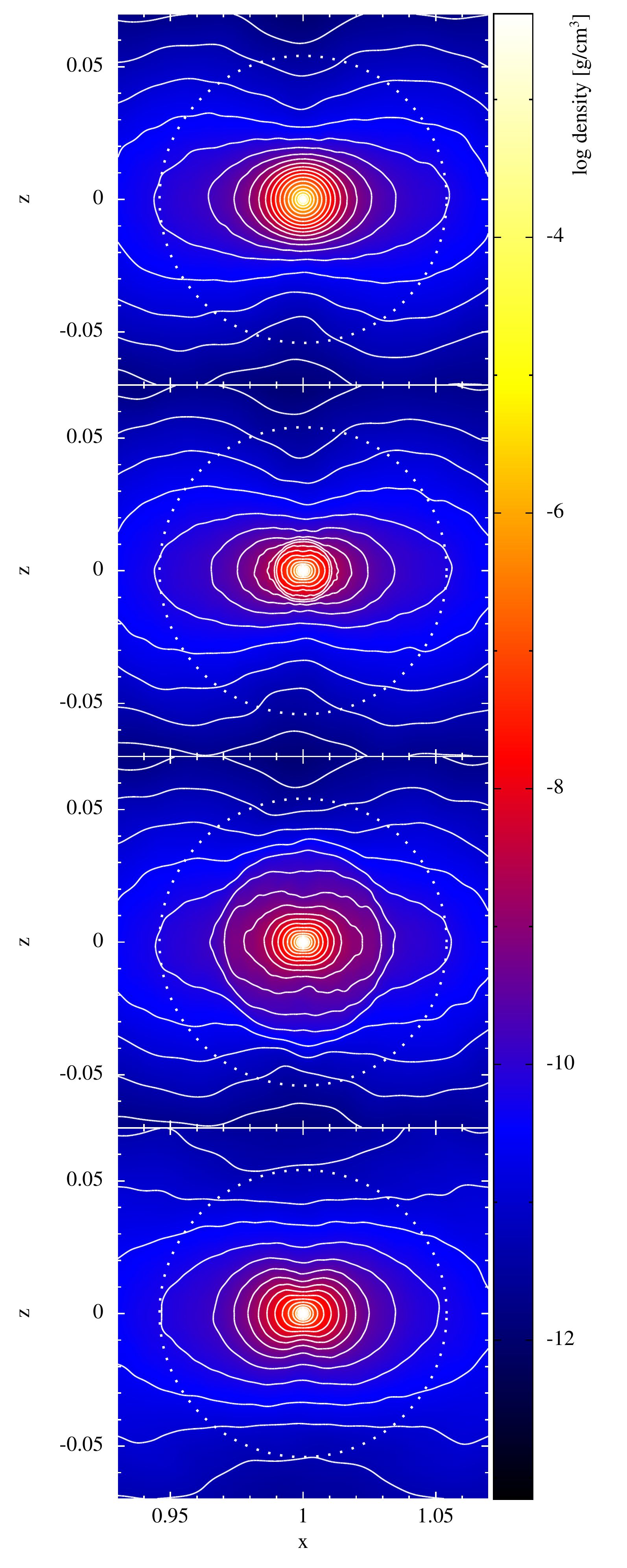}
}
\subfigure 
{
    \includegraphics[width=\columnwidth]{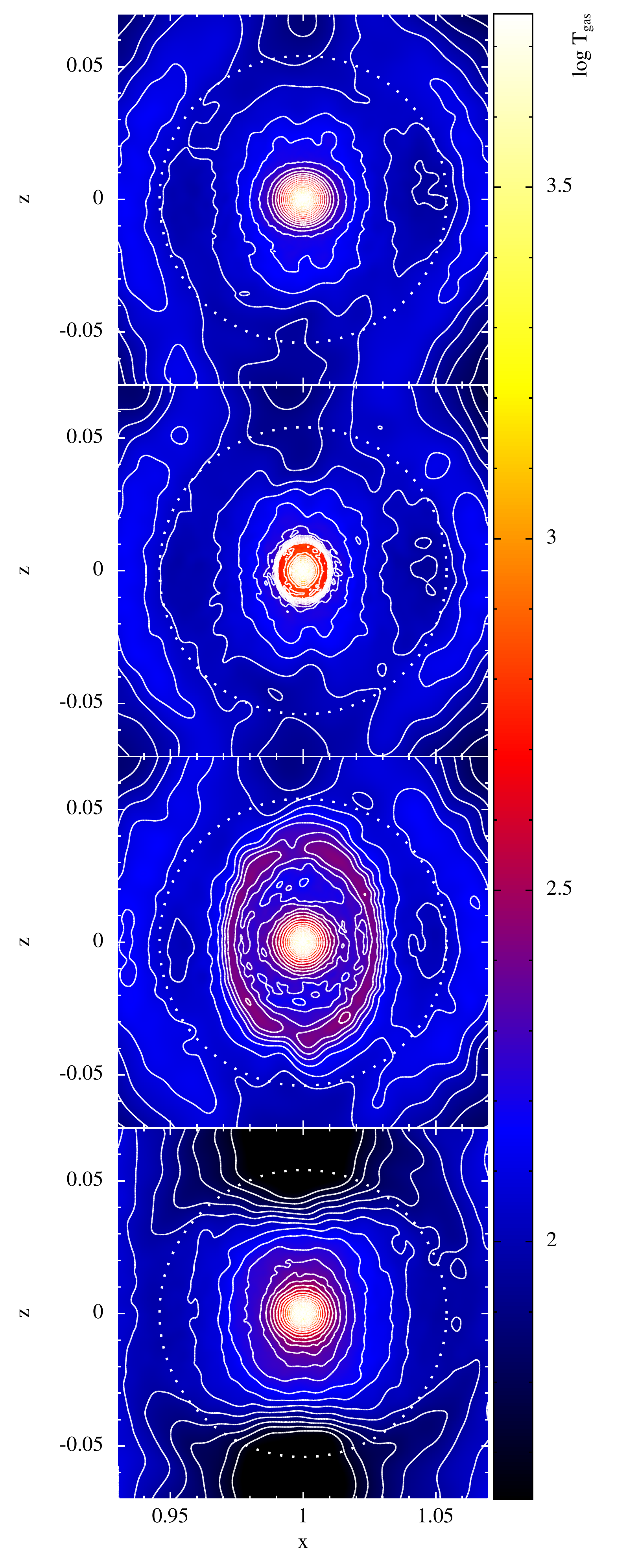}
}
\caption{Cross-section plots illustrating the envelope collapse about a 33~\earthmass \ ($\approx 0.1 {\rm M_{Jupiter}}$) core. The Hill radius is marked as a dotted line. The accreted mass (gas within \rhill) is $\approx 0.375$~\jupitermass \ prior to the collapse, reducing by $\approx 1.3$ per cent as the shock pushes material away from the core. The x and z axes are given in units of \rorbit. {\bf Left panels:} Cross sections in density illustrate the envelopes collapse, the shock propagation, and the formation of a circumplanetary disc. The central density increases by an order of magnitude from the top panel to the last. {\bf Right panels:} Temperature cross-sections at equivalent times to the density panels. The peak temperature increases by 2100K to 6800K from the first to the last panel. The protoplanetary disc scaleheight is $\approx 0.05$~\rorbit, thus there is little material obstructing the vertical propagation of the shock front. As a result, the shock propagates more easily in the vertical direction than through the denser midplane, yielding the non-spherical propagation most clearly shown in the third righthand panel.}
\label{fig:zdens}
\end{figure*}

The process of envelope collapse is presented in Fig.~\ref{fig:zdens}, the left hand panels of which shows cross-sections in density in the Z-X plane through the centre of the planet, from a time preceding the collapse in the uppermost panel, and at various stages during the collapse in subsequent panels. In the first panel the density contours are near spherical at small radii, elliptical at 0.5~\rhill, and pinched in towards the planet's poles at large radii. The second panel illustrates the strong shock propagating outwards from near the planet's core, which at this point has not altered the gas structure near the Hill radius (marked with a dotted line). However, it is possible to see that the collapse has been greater at the poles, deforming the previously elliptical contours. The third panel demonstrates the continued propagation of the shock, which is elongated along the vertical axis due to the lower density of material above and below the plane of the disc, which allows the shock to propagate more rapidly in this direction. Contours which had been pinched in at the planet's poles are forced outwards as the shock passes, but as can be seen, inside they are already resuming their pinched structure. The bottom panel illustrates the resettled state of the protoplanet and its surroundings. The pinching at the poles, which in the top panel was found beyond 0.5~\rhill, now extends down to $\approx 0.1$~\rhill, and the surrounding medium now forms a circumplanetary disc, having previously existed as an ellipsoidal envelope. The central density of the protoplanet, that is the gas density near the core, was $\approx 5 \times 10^{-4}$~\density \ in the top panel, and has increased to $\approx 5.5 \times 10^{-3} \mathdensity$ \ in the structure shown in the bottom panel following the collapse.

Returning to the formation of a circumplanetary disc, Fig.~\ref{fig:circum} illustrates the midplane and vertical density structures about the protoplanet, emphasising the altered state of the envelope. The divergence from a spherically-symmetric density distribution, defined here as a difference of greater than 10 per cent between the midplane and vertical distributions, occurs at a radius 5 times smaller in the post collapse state than in the pre-collapse state. This illustrates that the envelope undergoes a more significant change in its structure vertically, than it does in the plane of the disc, and is similar to the results seen in \cite{AylBat2009}. This was investigated further in \cite{AylBat2009a} where it was found that circumplanetary discs formed around massive protoplanets, and that these discs tended to be thick, with dimensionless scale heights generally larger than 0.2. We measure the circumplanetary disc scale height by taking radial bins within the region with $r < \mrhill/3$ (measured from the protoplanet).  In each radial bin, we take the SPH particle densities (from all azimuthal angles) and fit Gaussian profiles to the resulting vertical density distributions. A Levenberg--Marquardt algorithm is used to perform the fit, allowing the scale height to vary. This gives a measure of disc scale height versus radius, which in this case yields values for $H/r$ that increase from 0.4 to 0.5 over the radial range of 0.05$-$0.33~\rhill \ over which the disc extends. This radial extent of the circumplanetary disc can be seen in Fig.~\ref{fig:angmom} in which the disc edge is taken to be at the location where the peak of the specific angular momentum occurs.  Beyond this radius, the specific angular momentum decreases with radius since, in the frame rotating with the protoplanet, the material in the circumstellar disc is counter-rotating relative to the gas captured by the protoplanet. { Over the radial range 0.07$-$0.3~\rhill \ the specific angular momentum is on average $0.65$ of the Keplerian values (marked by the dashed line). The displacement is due to the pressure support within the thick circumplanetary disc, and the degree of this displacement can be used to approximately calculate the disc scale height as another check on the values measured above. Using the ratio of the specific angular momentum ($j$) to the Keplerian value ($j_k$) we obtain a scale height of 0.55. This is obtained using equation~\ref{eq:sheight} (see appendices B \& C of \citealt{LaiGonMad2012}), where we have used values of $1/2$ and $7/10$ for the surface density and temperature exponents ($p$ and $q$) for the circumplanetary disc, typical values from \cite{AylBat2009b}; the calculated scale height is not enormously sensitive to variations in these values within a reasonable range. 

\begin{equation}
\label{eq:sheight}
\frac{H}{r} = \sqrt{ \frac{2 (1 - j/j_k)}{ p + q/2 + 3/2}}
\end{equation}

\noindent This resulting scale height is somewhat larger than that which we directly measured, but has been calculated assuming a vertically isothermal disc and a lack of self-gravity, and is thus only approximate. The specific angular momentum distribution further supports our assertion that a circumplanetary disc has formed as a result of the envelope collapse.}

\begin{figure}
\centering
\includegraphics[width=\columnwidth]{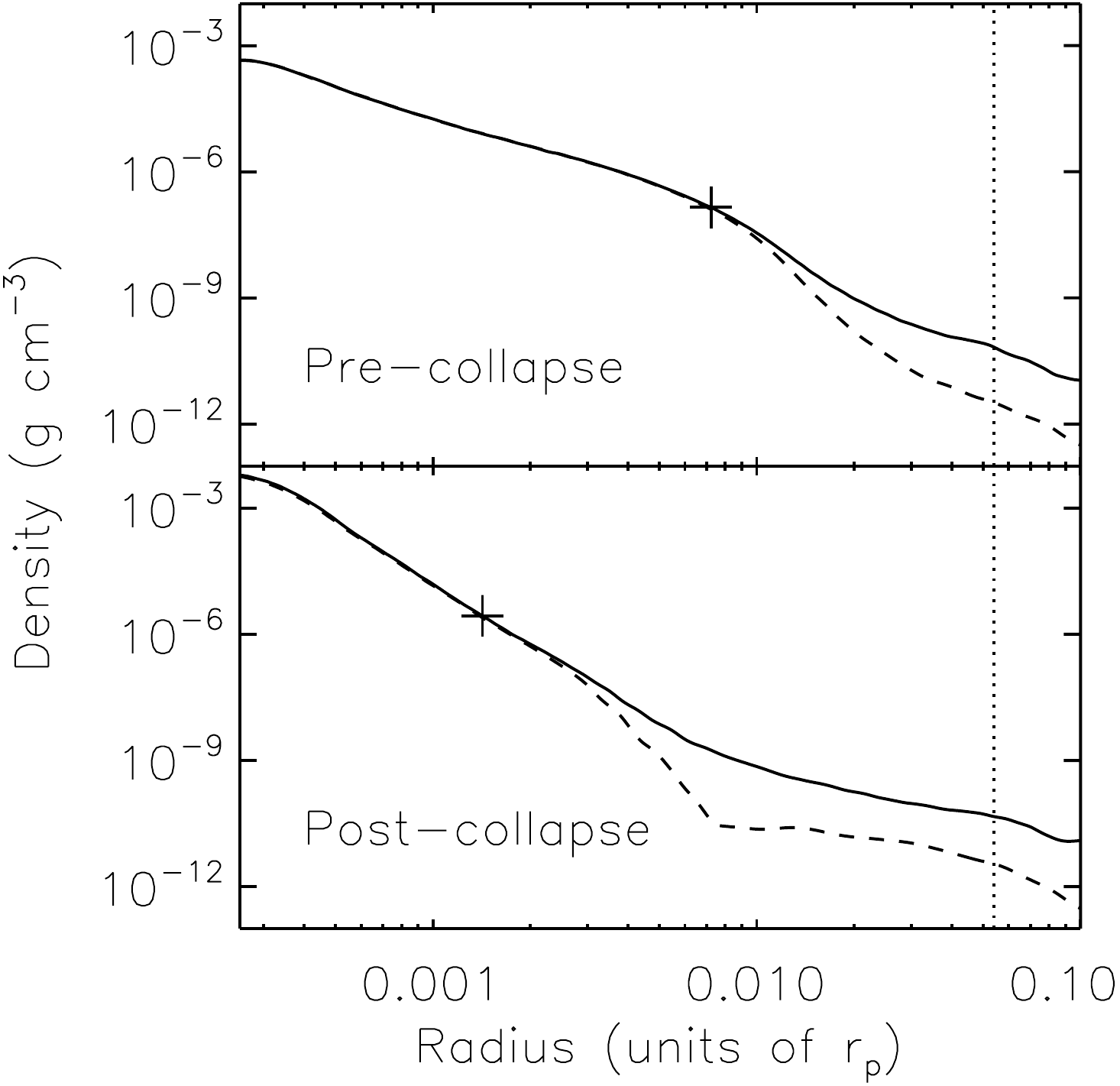}
\caption{The density distribution along the x-axis (solid lines) and along the z-axis (dashed lines) through the protoplanet before (top panel) and after (bottom panel) collapse. The plus symbols mark the radius and density at which the midplane and vertical density distributions differ by more than 10 per cent. The vertical dotted line marks the protoplanet's Hill radius. The density distribution is modified significantly in both the midplane and vertically following collapse, but the point at which these distributions diverge moves inwards by a factor of 5 in radius.}
\label{fig:circum}
\end{figure}

\begin{figure}
\centering
\includegraphics[width=\columnwidth]{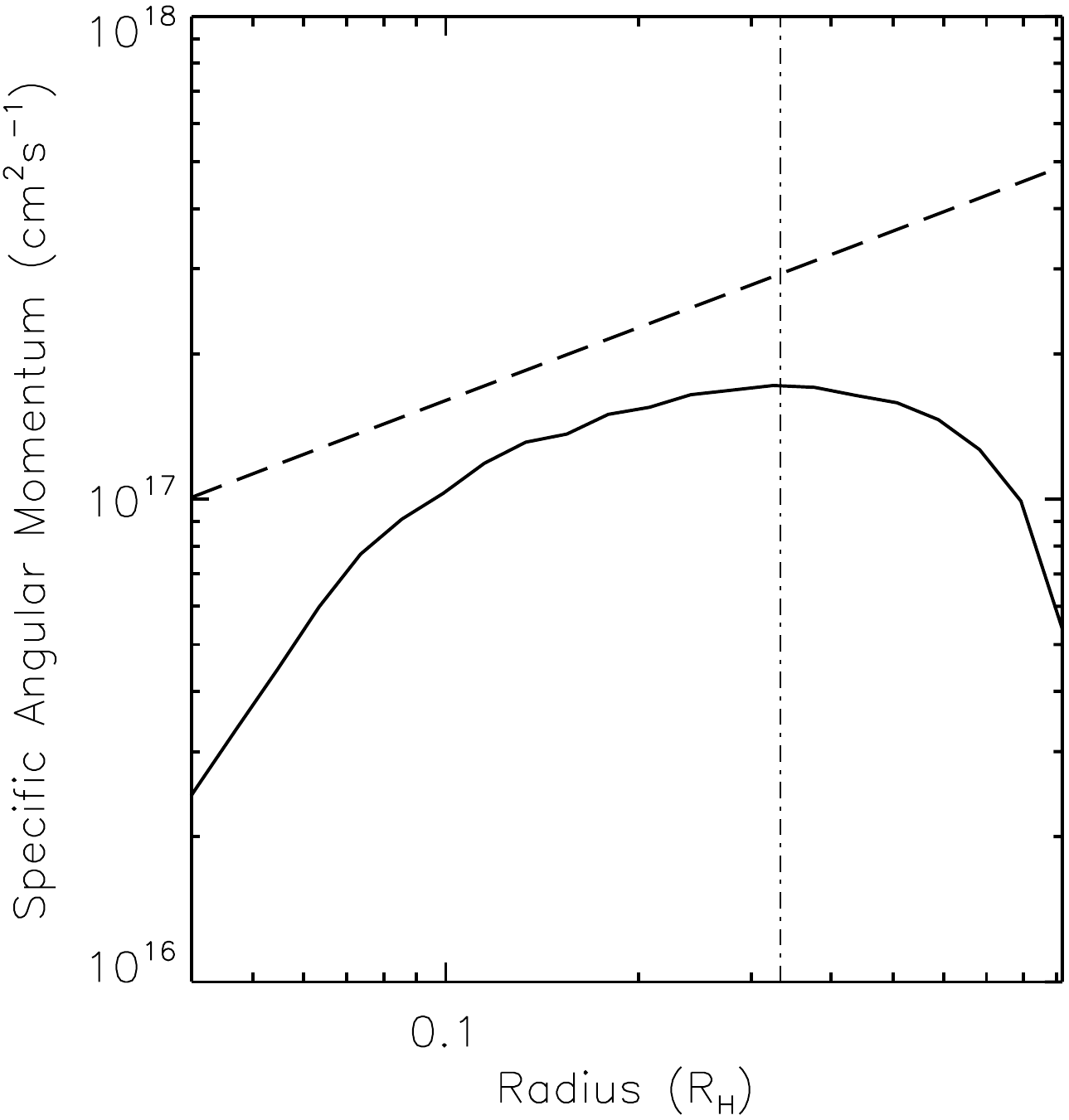}
\caption{ The specific angular momentum of the gas surrounding the protoplanet that comprises the circumplanetary disc. The vertical dot-dashed line marks \rhill/3, the analytically expected edge of the disc, and this matches the measured turnover very well. The dashed line marks the Keplerian orbital velocity based on the mass within the associated radius. Pressure support within the disc means that a sub-Keplerian orbital velocity is expected, though with a similar gradient if the disc is rotating about the planet. This gradient matches reasonably between 0.07$-$0.3~\rhill.}
\label{fig:angmom}
\end{figure}

\begin{figure}
\centering
\includegraphics[width=\columnwidth]{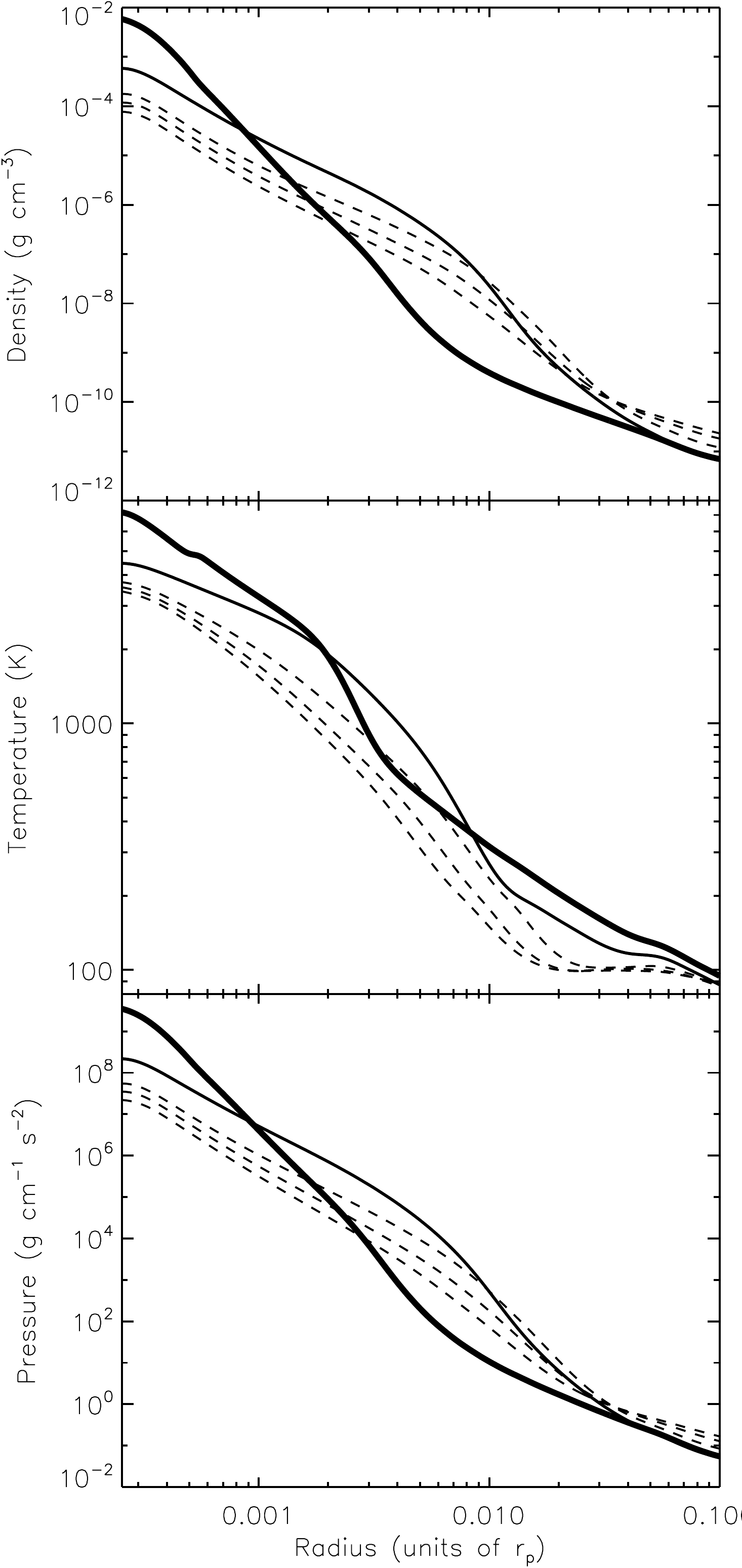}
\caption{Spherically averaged density (upper panel), temperature (middle panel), and pressure (lower panel) distributions about the protoplanet. The dashed lines are at times of 50, 100, and 150 orbits, their order ascending up the left hand axis. The solid lines show the pre-collapse state (thinner line), and the post-collapse state (thicker line). The pre-collapse state is equivalent to that in Fig.~\ref{fig:cmass}, whilst the post-collapse state is instead taken at the very end of the calculation in this case, when accretion has resumed its pre-collapse rate.}
\label{fig:denstemp}
\end{figure}

The right hand panels of Fig.~\ref{fig:zdens} shows the temperature structure in a Z-X slice at equivalent times to those shown in the density panels to the left, allowing us to see the temperature evolution during the collapse and shock propagation. A hot front associated with the shock can be seen expanding away from the core with time, and the vertical elongation of the shock can be clearly seen in the third panel. Near the protoplanet's core, the peak temperature has increased by 2100K to more than 6800K in the post-collapse state shown in the final panel, and continues to increase for the remainder of the calculation when accretion has recommenced.

\begin{figure}
\centering
\includegraphics[width=\columnwidth]{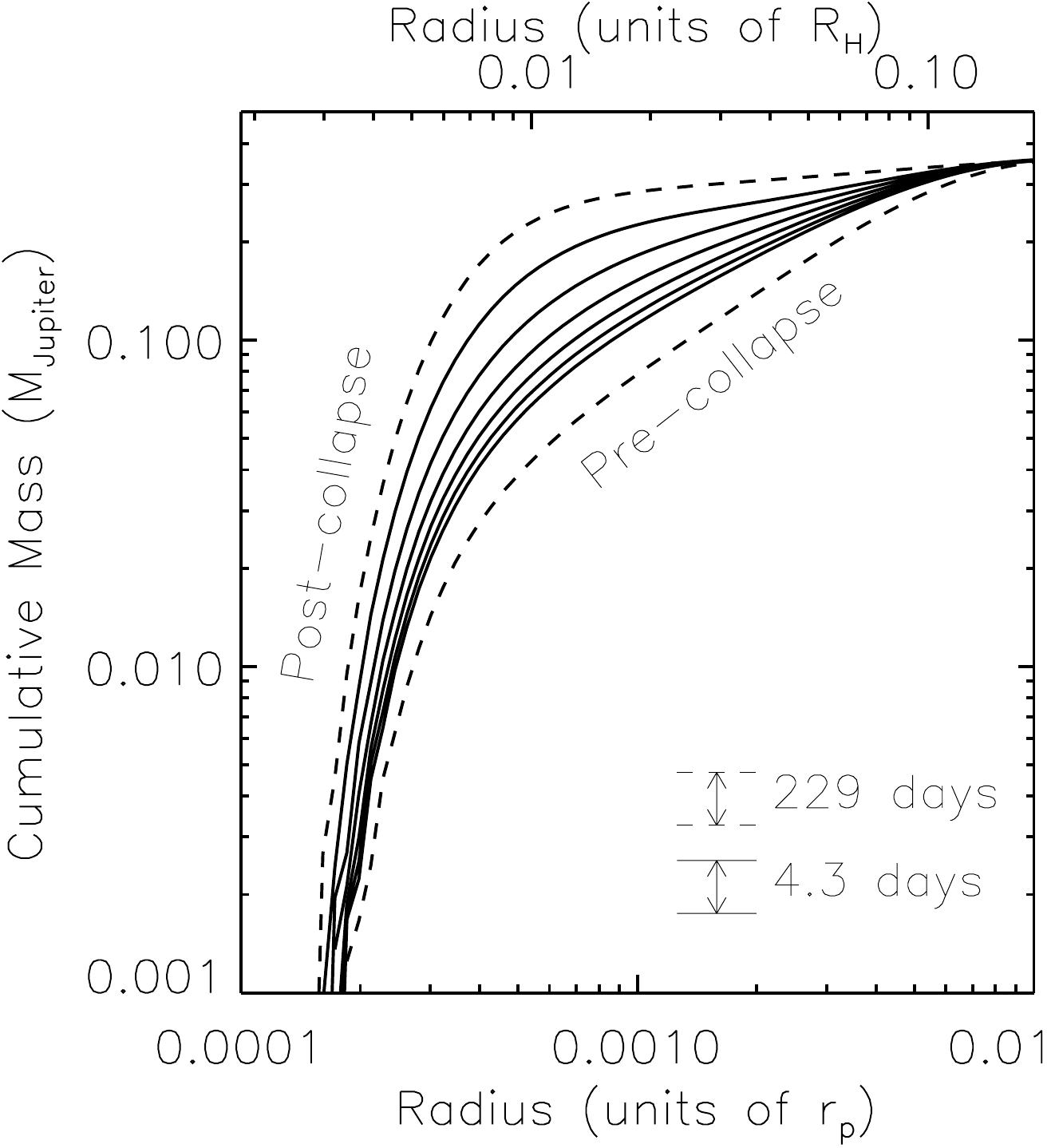}
\caption{Cumulative mass distribution calculated outwards from the core of Model J, where $M_{\rm acc} = 0.375$~\jupitermass. The mass distribution is shown over times ranging from just before the collapse, to after the structure stabilises. The collapse proceeds very rapidly, as shown by the solid lines with cover a period of less than 22 days from first (lowest) to last (highest).}
\label{fig:cmass}
\end{figure}

Fig.~\ref{fig:denstemp} shows the spherically averaged radial density and temperature profiles running outwards from the protoplanetary core at a number of points in the calculation. The change between 50 and 150 orbits (the dashed lines) is relatively small, increasing as would be expected for a growing protoplanet. Arriving at the pre-collapse state (thin solid line) at around 190 orbits, at which point the accretion rate is at its maximum, the density and temperature maxima have increased more over 40 orbits than in the preceding 100 orbits. Moreover, the temperature structure shows a marked change in its form. However the most significant changes occur during the collapse and 5 further orbits of evolution (thick solid line). The envelope's collapse occurs very rapidly, and only stops when the new structure of the envelope is able to reestablish hydrostatic equilibrium. For this to occur the density structure changes to deliver a much steeper gradient away from the planet's solid surface. This leads to a much steeper pressure gradient in this region, as shown in the bottom panel of Fig.~\ref{fig:denstemp}, eventually satisfying the requirement $\nabla P = -\rho \nabla \phi$, where $\phi = {\rm G}M(r)/r$. As such, the envelope is able to resume steady accretion as the structure is able to bear the increasing weight; as mentioned previously, accretion resumes at the pre-collapse rate.

Fig.~\ref{fig:cmass} depicts the changing distribution of mass in the inner envelope from its pre-collapse state, to its post-collapse state. The total time span shown is 229 days, the time between pre and post collapse within the inner envelope, with these states marked using dashed lines. However, the most significant changes occur over less than 22 days at this scale, and this period is broken down in Fig.~\ref{fig:cmass} into 4.3 day increments which are marked with solid lines. The Hill radius just prior to the collapse is equal to 0.054~\rorbit, which taking \cite{LisHubDAnBod2009}'s estimate of a $\sim 0.25$~\rhill \ envelope radius, gives a size of 0.0135~\rorbit. This radius is in reasonable agreement with the region over which the mass is significantly redistributed during the collapse, which can be seen in Fig.~\ref{fig:cmass} to be 0.01~\rorbit \ ($\approx 0.2$~\rhill).

\begin{figure}
\centering
\includegraphics[width=\columnwidth]{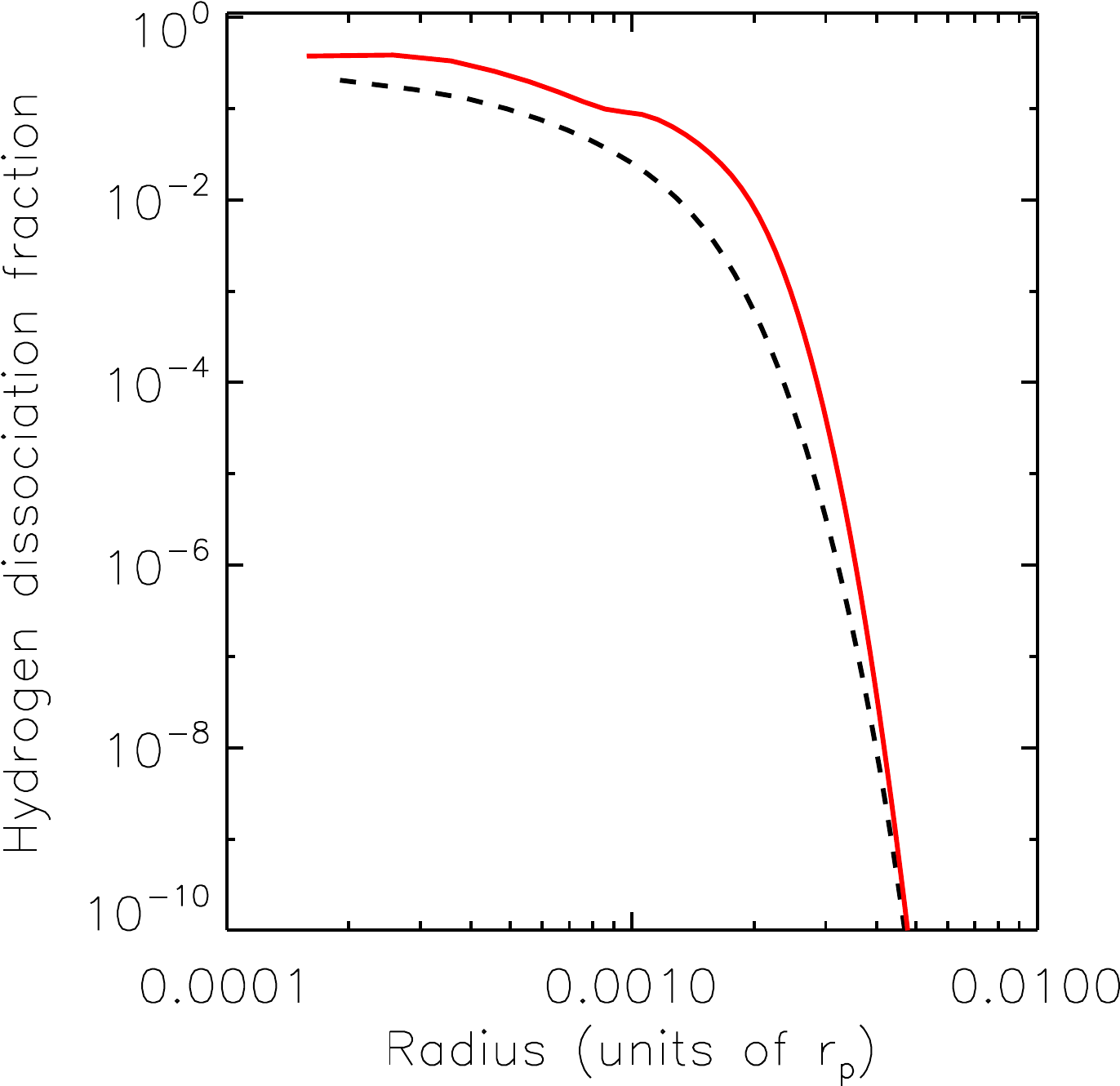}
\caption{Fraction of atomic hydrogen versus radius within the inner region of the protoplanetary envelope before it collapses (dashed line), and after it collapses (solid line). The higher temperatures that develop within the deep envelope when it collapses lead to a higher dissociation fraction of molecular hydrogen, whilst this process of dissociation will absorb energy, reducing the maximum temperature that is achieved. Prior to collapse the fraction of atomic hydrogen peaks at 0.33, whilst in the immediate aftermath it is as high as 0.48.}
\label{fig:diss}
\end{figure}

Collapse of the protoplanetary envelope leads to a substantial increase in the temperature near the protoplanetary core, as discussed above. A result of this temperature rise is an increase in the dissociation fraction of molecular hydrogen about the core, and an enlargement of the region within which hydrogen is substantially dissociated; this can be seen in Fig.~\ref{fig:diss}. It is not the dissociation that triggers the collapse of the envelope, despite the process acting as an energy sink; the capacity of dissociation to absorb energy does lead to a lower final temperature within the envelope than might otherwise have been achieved.

\begin{figure*}
\centering
\includegraphics[width=\textwidth]{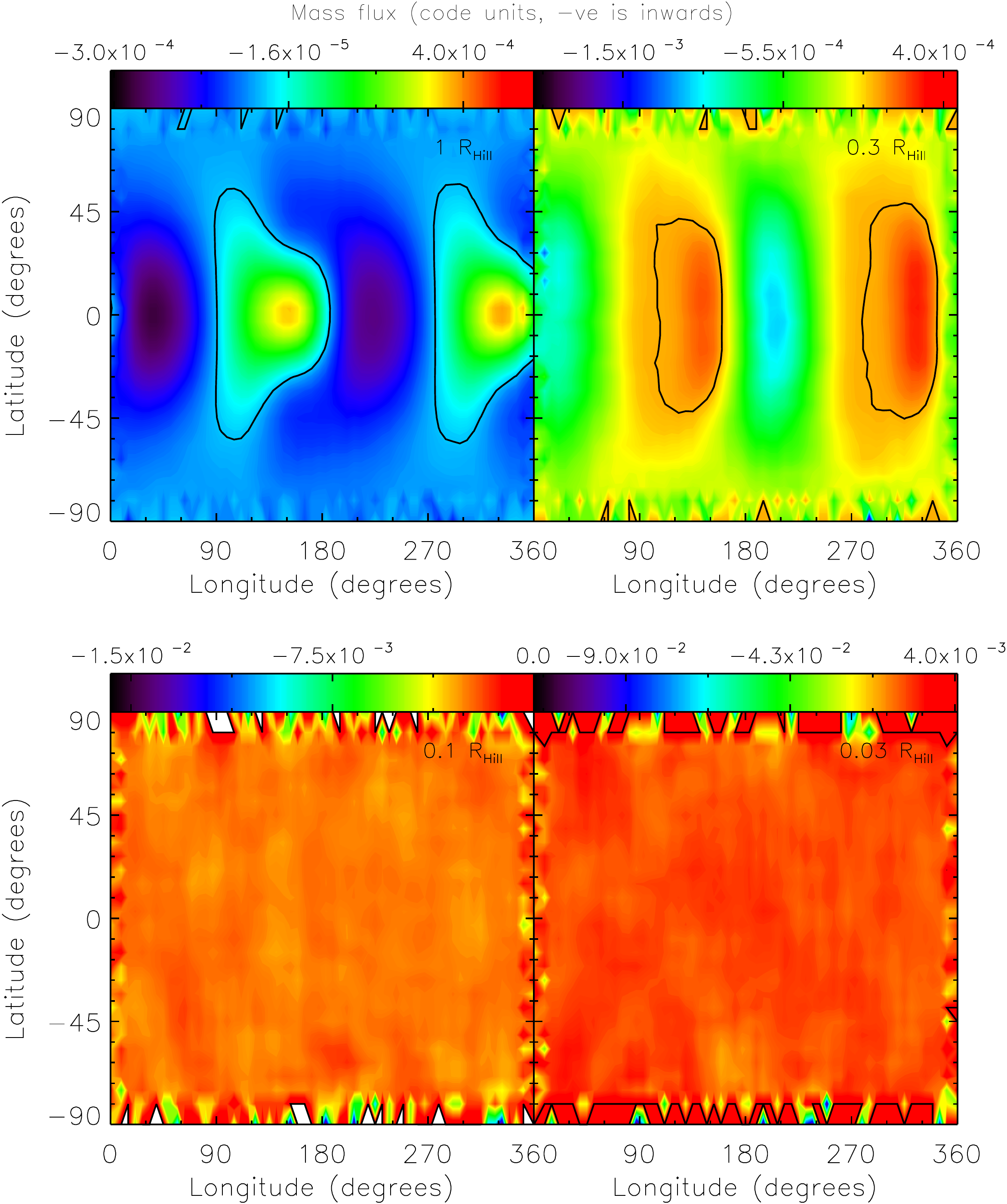}
\caption{Mass flux through shells of various radii (marked in panels) surrounding the protoplanet core in Model J prior to its envelope collapse. At one Hill radius, the flow takes on a form that is similar to that seen in Fig.~5 of \protect \cite{TanOhtMac2012}, and reflects the combined effect of horseshoe orbits and bent flow lines passing the protoplanet. Material both enters and leaves the sphere predominantly at the midplane where densities are highest; the marked contour line denotes a flux of zero such that the area within the contour marks the outflow. For ease of comparison with \protect \cite{TanOhtMac2012} the solar and anti-solar points are at longitudes of 0 and 180 degrees respectively, and inflowing material is shown by a negative flux. At the smallest radii the flow is largely inwards across the shell, with an average flux of $-1 \times 10^{-2}$~code units ($-5.5 \times 10^{-5}~{\rm g \: cm^{-2} \:  s^{-1}}$).}
\label{fig:preflow}
\end{figure*}

\begin{figure*}
\centering
\includegraphics[width=\textwidth]{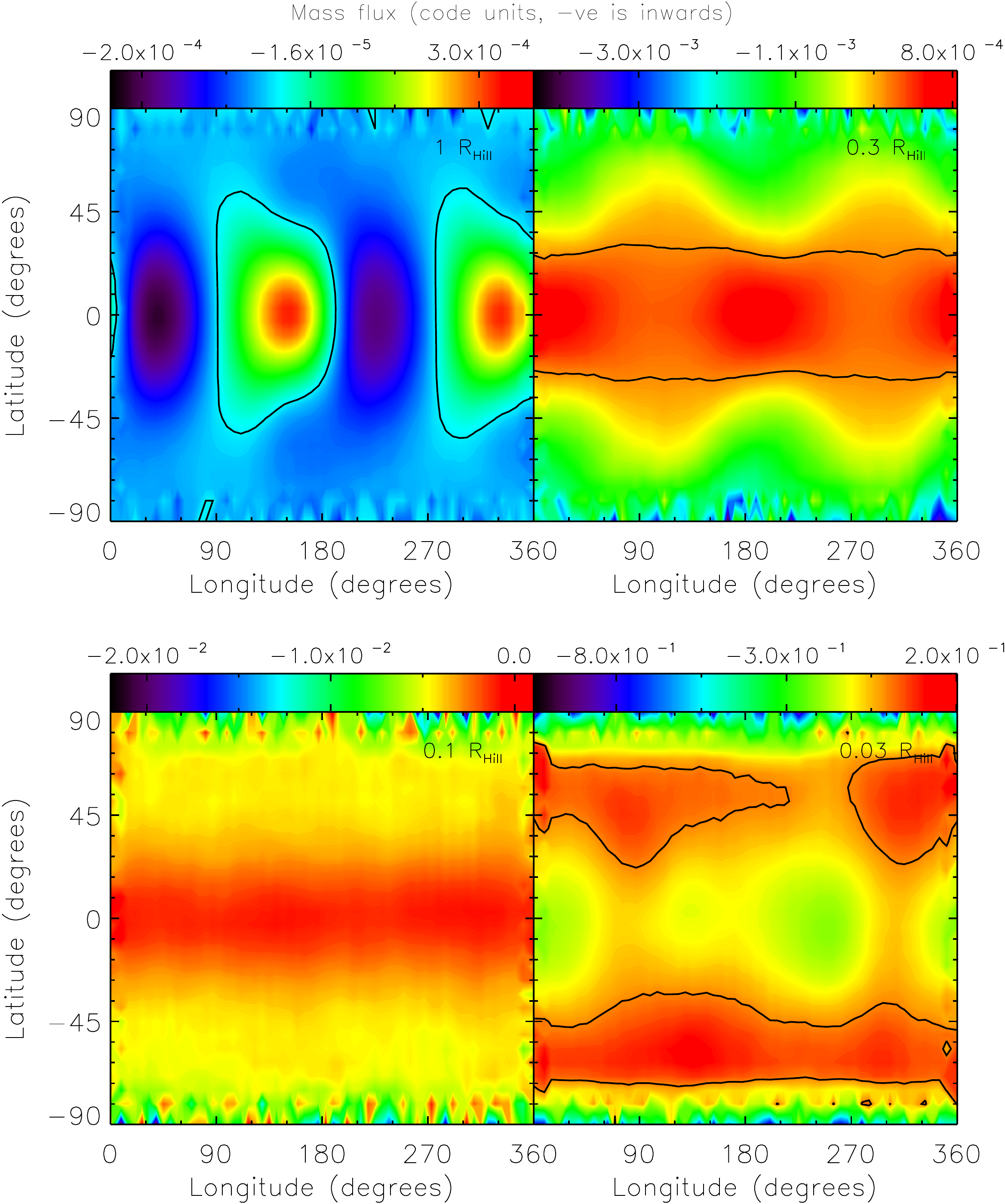}
\caption{Identical to Fig.~\ref{fig:preflow} except illustrating the gas flow after Model J undergoes its envelope collapse. The flow at one Hill radius is very similar to the pre-collapse case, however at radii between $0.1 - 0.3$~\rhill \ the outflow is concentrated along the midplane, whilst inflow occurs at higher latitudes, reflecting the vertical accretion seen by \protect \cite{MacKokInuMat2008}. At the smallest radii, there is evidence of a gas turn over, where material is flowing in along the midplane and at the poles, and out in a range of moderate to high latitudes (see Section~\ref{sec:convection}).}
\label{fig:postflow}
\end{figure*}

\subsection{Accretion flow}
\label{sec:flow}

Whilst we find that a protoplanet envelope increases in mass throughout its evolution, excepting a brief period following a dynamical collapse (seen at radii of 1, 0.3, and 0.1~\rhill \ in Fig.~\ref{fig:racc}), it is not obvious that this accretion is a spherically symmetric process. \cite{MacKokInuMat2008} and \cite{TanOhtMac2012} have performed three-dimensional calculations of protoplanet growth, and find that gas flows outwards along the midplane from a growing protoplanet, such that accreted material must be delivered vertically. In \cite{AylBat2009a} we found that mass predominantly entered the Hill sphere along the midplane, but this analysis was simplistic in that it only considered inflow, failing to consider the possibility of outflow, and we worked solely with the azimuthally integrated values. Here we make a more thorough assessment of the mass flow, and are able to look at this flow in a large extended envelope that has not undergone collapse, as well as in the dense envelope formed subsequent to such a collapse. It is the latter case which most closely resembles the principle model of \cite{TanOhtMac2012} for a high mass protoplanet.

Before and after the envelope collapse, we see material flowing in and out at the Hill radius in an alternating pattern, as can be seen in the first panels of Figs.~\ref{fig:preflow} \& \ref{fig:postflow}, where negative values correspond to inflow, and the solid contour line denotes a flux of zero. This flow is easily explained as the passage of gas passing the planet and being deflected by the spiral shocks that form due to the gravitational perturbation provided by the protoplanet. Another contribution comes from gas following horseshoe orbits which also enter and leave the Hill sphere at broadly similar longitudes about the planet. Fig.~\ref{fig:vfield} illustrates the vector field that results in the mass flow observed at the Hill radius, which is marked in this figure with a dashed line. The data presented in Figs.~\ref{fig:preflow} \& \ref{fig:postflow} was constructed by considering the flow over a period of 4 orbits preceding and following the collapse respectively. The stability of the gas flow about the protoplanet over these periods leads to the regular pattern that is seen at the Hill radius in both figures. In the pre-collapse case the pattern persists down to 0.3~\rhill \ as the second panel of Fig.~\ref{fig:preflow} illustrates. At smaller radii the flow is inwards across the spherical shell, with a larger flux at smaller radii as is expected for a consistent mass flowing across a decreased surface area. In the pre-collapse state the mass flow at small radii does not appear to possess any latitudinal dependence, rather it flows in almost spherically symmetrically. Note that the figures are noisey at the poles as a result of the spherical polar grid used to calculate the flux, which leads to very small bins at high latitudes.

In the post-collapse case, shown in Fig.~\ref{fig:postflow}, the mass flux at the Hill sphere is little changed from the pre collapse case, and in both cases it is reminiscent of the structures seen in Fig.~5 of \cite{TanOhtMac2012}; note that we are plotting a time average flux, whilst \citeauthor{TanOhtMac2012} plot an instantaneous flux ($\rho v_{\rm r}$). Unlike \cite{TanOhtMac2012}, who see these alternating structures persisting down to very small radii, our model has already lost any sign of the in-out flow pattern at a radius of 0.3~\rhill. Instead, at 0.3~\rhill \ we find an outflow along the circumplanetary disc midplane at every longitude, though this outflow shows a slightly alternating magnitude, with more significant outflow at longitudes of $\sim 20$ and 200 degrees. This outflow originates at a radius of $\approx 0.17$~\rhill, which is around half the radius of the circumplanetary disc, inside of this radius the flow is inwards. Meanwhile at 0.3~\rhill, mass is flowing inwards at higher latitudes. The relative fluxes are such that the net flux at each radius is negative, enabling the protoplanet to continue to grow. This growth is corroborated by Fig.~\ref{fig:racc}, which shows the mass evolution of Model J within the 4 radii considered here, and indicates that the mass consistently increases within 0.3~\rhill \ (dashed line) once the post-collapse state is achieved. Following the envelope's collapse, there is more discernible structure to the mass flow at the smaller radii. At 0.1~\rhill \ the inflow along the circumplanetary disc midplane is not as significant as the flow at high latitudes, as is shown in the third panel of Fig.~\ref{fig:postflow}. This panel marks a clear divide in the motion of gas at and around a radius of 0.1~\rhill, where the flow is only found to be inwards, signalling that material within this radius is truly bound to the protoplanet. This is of particular interest because of the gas flow found at 0.03~\rhill, and shown in the final panel of Fig.~\ref{fig:postflow}. At this small radius there are significant flows of material, pushing outwards at intermediate latitudes ($\sim 45 - 70$~degrees), and pouring inwards again along the midplane. This is indicative of significant circulation of the bound material below 0.1~\rhill, and will be discussed further in Section~\ref{sec:convection}.

\section{Discussion}
\label{sec:discussion}

\subsection{Hydrodynamic collapse}

From the earliest suggestion of \cite{PerCam1974} it has been thought that a giant planet might form through the hydrodynamic collapse of a gaseous envelope onto a solid core which caused it to assemble. This was followed by numerous models that effectively sought for hydrostatic solutions to various combinations of properties to establish when such a collapse might occur \citep{MizNakHay1978,Miz1980,Sas1989}. The first models that attempted to model giant planet growth from the initial core formation, through to the envelope growth were performed by \cite{BodPol1986}. These models revealed that a protoplanetary envelope would gradually contract as the planet grew, leading to a quasi-static contraction beyond previously calculated values for the critical mass, as long as the envelope did not effectively detach from the protoplanetary disc (that is there was a sufficiently rapid supply of material from the latter to the former). Under these conditions, their was no evidence to suggest that the hydrostatic balance should reach some limit beyond which a collapse was inevitable, and later semi-analytic models originating from these earlier works, such as \cite{PolHubBodLis1996}, \cite{HubBodLis2005}, and \cite{LisHubDAnBod2009}, suggest no need for a dynamic collapse. Our Model J follows a pattern of stable growth for the vast majority of its history, though not evidently undergoing any significant contraction, and resumes this pattern of growth subsequent to its envelope collapse.

\begin{figure}
\centering
\includegraphics[width=\columnwidth]{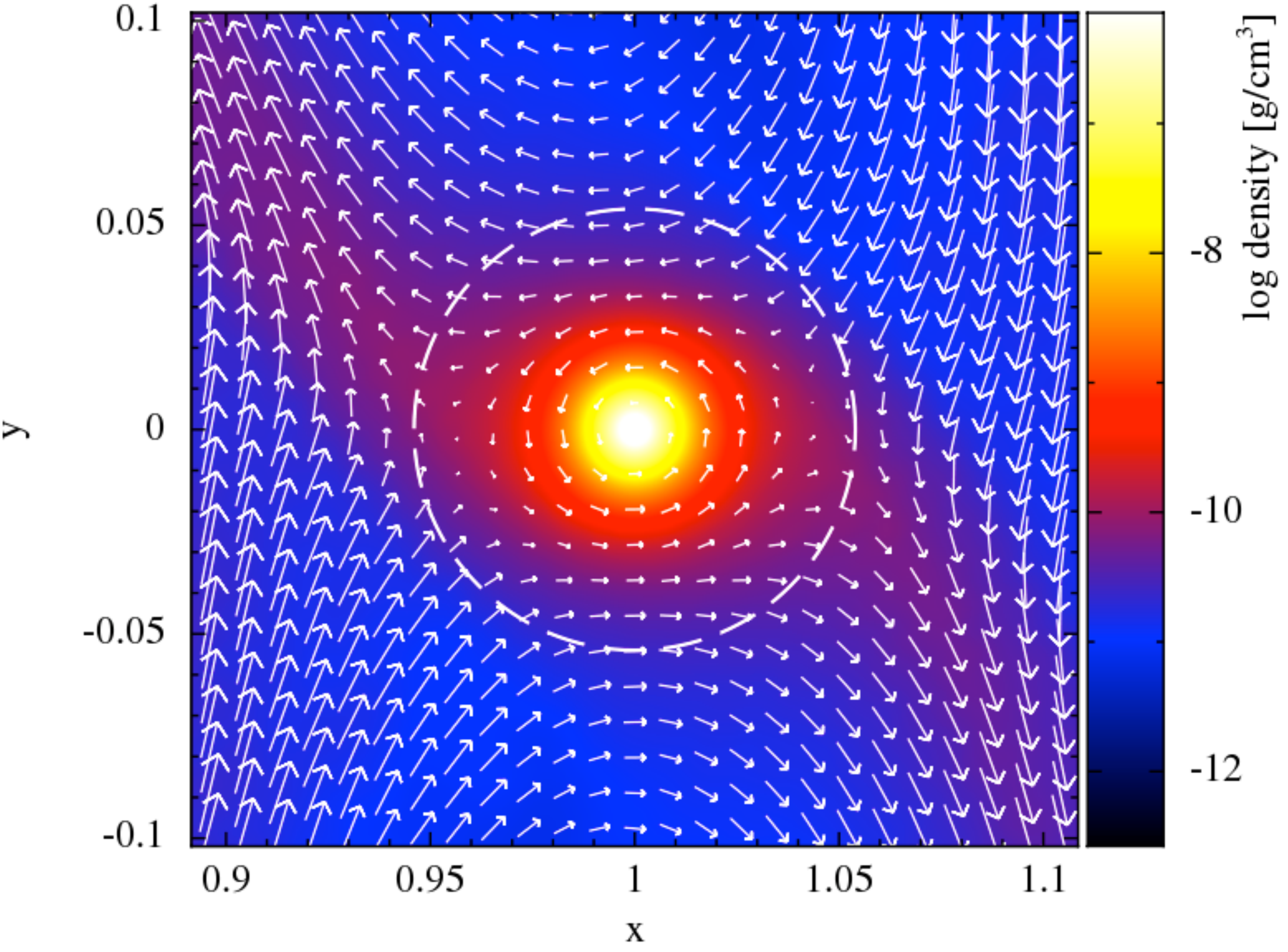}
\caption{Velocity vectors in the plane of the disc, illustrating the gas flow about the Hill radius (marked with a dashed line) that leads to the alternating pattern of in and out flow seen in the first panels of the mass flux plots shown in Figs.~\ref{fig:preflow} \& \ref{fig:postflow}. This vector field is plotted for a time preceding the envelope collapse, but a very similar field exists after the shock associated with the collapse has passed out of the region.}
\label{fig:vfield}
\end{figure}

The models presented in this article are performed using a three-dimensional hydrodynamics code that include self-gravity, and radiative transfer, but which omits the core formation phase, and the deposition of energy due to planetesimal accretion that are included in the semi-analytic works discussed. However, at the time of interest around the envelope collapse, the energy release is utterly dominated by the contraction of the gaseous envelope, such that solids accretion energy may be regarded as negligible. The metallicity of the envelope might be significantly modified by the ablation and evaporation of grains that have been accreted over the planets history, and we make no attempt to account for this. The opacity in Model J is reduced by a fixed factor of $10^{3}$, { at the lowest end of the range suggested by \cite{MovBodPodLis2010}, who found such opacities due to grain settling and coagulation in regions of the envelope.}

It is difficult to disentangle the causes and effects of the very rapid collapse we find, for example the surge in temperature leads to a higher fraction of dissociated hydrogen. However, as stated in Section~\ref{sec:collapse}, it does not appear that the dissociation of molecular hydrogen acts to trigger the collapse, as the fraction remains steady in the preceding period. There is also no evidence of the Kappa-mechanism acting within the protoplanetary atmosphere in our model, as was found by \cite{Wuc1991} to cause a dynamic collapse. \citeauthor{Wuc1991} also found that this collapse led to a significant ejection of material from the protoplanet, whilst we find only a small drop of $\approx 1.3$ per cent in the mass within the Hill radius, and a rapid increase at and below 0.1~\rhill \ as the protoplanet structure shifts to its new state.

\begin{figure}
\centering
\includegraphics[width=\columnwidth]{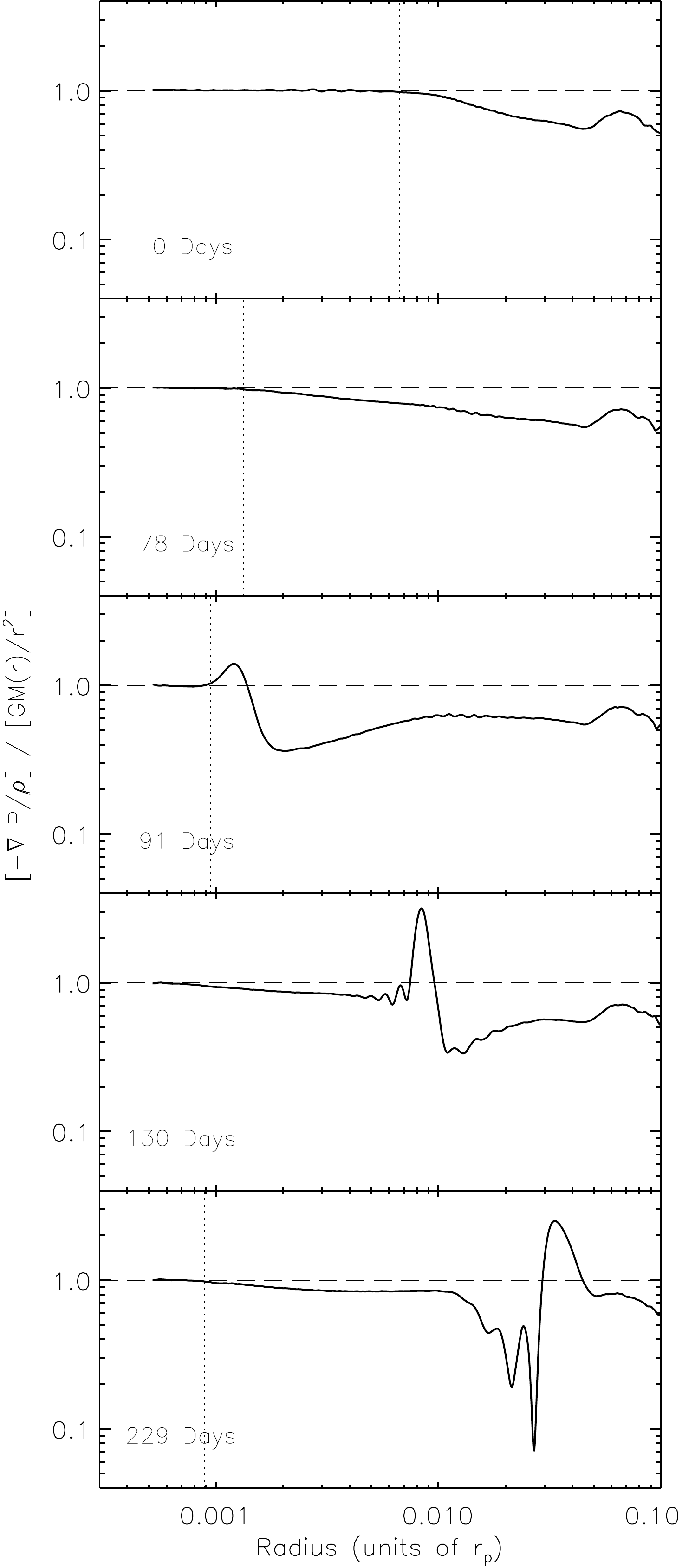}
\caption{Each panel shows the ratio of the pressure force to the gravitational force in the envelope and beyond, with the first and last panels corresponding in time to the pre and post-collapse cumulative mass distributions shown in Fig.~\ref{fig:cmass}. A ratio of one indicates that the material is in hydrostatic equilibrium, which before the envelope collapse applies to a region out to 0.0067~\rorbit \ (0.12~\rhill, top panel, marked with vertical dotted line), but which post-collapse reaches out to just 0.0009~\rorbit \ (0.016~\rhill). By the final panel the shock wave has cleared the inner 0.01~\rorbit, leaving the environment internal to this in a new hydrostatic equilibrium, whilst its mass continues to increase, as shown in Fig.~\ref{fig:racc}.}
\label{fig:hydrocollapse}
\end{figure}

It appears that our Model J protoplanet reaches the hydrostatic limit for its formative structure, and that the internal pressure gradient can no longer accommodate the addition of mass by a small adjustment. This is illustrated in Fig.~\ref{fig:hydrocollapse} which shows the ratio of the pressure force to the gravitational force against radius at a number of stages of the envelope collapse. From an initial state, in which the protoplanetary atmosphere is in hydrostatic equilibrium out to a radius of 0.0067~\rorbit \ (0.12~\rhill), the atmosphere rapidly begins to restructure, pushing the hydrostatic region down to a radius of $\approx 0.0013$~\rorbit \ (0.025~\rhill). This initial stage leaves the form of the graph otherwise relatively unchanged, but as the central concentration of mass continues, a shock begins to form as the pressure near the core surges. The maximum pressure within 0.001~\rorbit \ has increased by an order of magnitude between the first panel and the third, leading to a somewhat steeper gradient over this region. However, at this point in time the gradient between $0.001 - 0.002$~\rorbit \ steepens much more rapidly, forming a shock front, and this front marks the new radial limit of hydrostatic equilibrium as can be seen in the third panel. The subsequent panel shows the shock propagating outwards, whilst the hydrostatic core shrinks a little. By the final panel the inner envelope has resettled and the structure has stabilised, whilst the shock's propagation continues outwards, eventually moving beyond the limits of the modelled region.

It is possible that this envelope collapse only occurs due to the high accretion rates achieved due to our selected disc conditions. The low opacity assumed promotes very rapid planet growth, and it may be this rapidity that prevents the envelope from adjusting its structure more gradually to accommodate the increasing mass. As such, it may be that the hydrodynamical collapse of a planetary atmosphere can only occur if that planet if accreting very rapidly. Further models will be required to determine whether or not this is the case. We note however that the accretion rate measured in Model J, both just before and after the collapse, is still not as rapid as would be found using a locally-isothermal equation of state, despite the large reduction in opacity. \cite{LisHubDAnBod2009} present accretion rates in their fig.~3, where these rates were obtained by \cite{DAnKleHen2003} using three-dimensional hydrodynamical models with a locally-isothermal equation of state. Applying our disc conditions to their results yields an accretion rate of $8 \times 10^{-4} ~{\rm M_{Jupiter}~year^{-1}}$, which is twice the rate measured in our radiation hydrodynamics { models prior to collapse}.

{ At this juncture we note that the results given in \cite{LisHubDAnBod2009} show a viscosity dependence, where higher viscosities lead to more rapid accretion. In our calculations viscosity is not a constant, but is proportional to the spatial resolution of the SPH method. Thus, the viscosity is lower in regions of higher density, and these differences mean the above comparison is only approximate. Further, we note that in the absence of a protoplanet, the unperturbed circumstellar disc in our models has a viscosity of $\alpha \approx 4 \times 10^{-3}$, consistent with the fixed viscosity global models of \citet{BatLubOgiMil2003} that are used to inject gas at the boundaries of the disc section. It is these boundaries that determine the rate at which gas is supplied to the disc section in these local models. As such, once the disc is perturbed, the spatially varying viscosity of the SPH calculations leads to to an inconsistency with the global models, and so the boundaries. A further caveat arising from the boundary implementation is that the injected material comprises gas on both circulating and librating orbits \citep{LubSeiArt1999}, orbits that are modified as the gas passes through the modelled disc section. However, these modifications are lost when the gas leaves the section, and new gas is injected without these modifications, leading to a further inconsistency.}

\subsection{Atmospheric turn over}
\label{sec:convection}

As briefly mentioned in Section~\ref{sec:flow} in reference to the third panel of Fig.~\ref{fig:postflow}, there appears to be significant motion of the bound gas besides rotation in the plane of the disc. An apparent rolling motion is indicated by the flow through the surface at 0.03~\rhill, the gas unable to escape, but flowing out and in within the bound atmosphere. These flows appear to be convection cells in the deepest parts of the planetary envelope where the medium is extremely optically thick, and thus unable to cool readily by radiation. { Fig.~\ref{fig:convection} shows a region out to $0.07$~\rhill \ ($3.7 \times 10^{-3}$~\rorbit), at which radius the temperature is 675K, increasing to a maximum of 7400K near the core, demonstrating a significant thermal gradient against radius.} \cite{BodPol1986} found strata of convection within protoplanetary envelopes in their one-dimensional models, with a dense inner convection region that contained some 95 per cent of the envelope mass, and depending on the age and core size, multiple higher level convection zones were also found to form. However, it is likely that the convection in fully 3-D models that include rotation, is very different.

\begin{figure}
\centering
\includegraphics[width=\columnwidth]{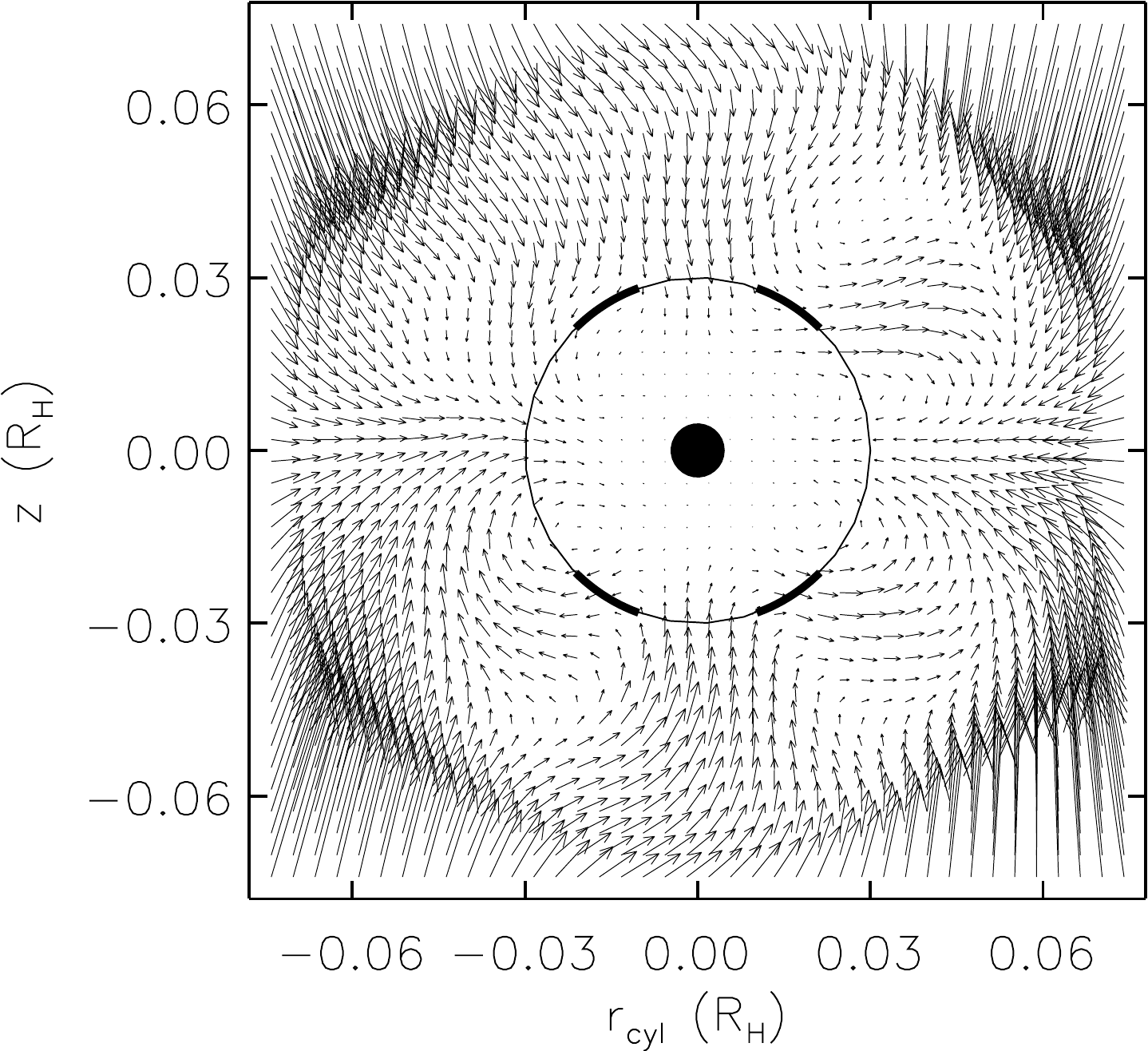}
\caption{A slice through Model J after the envelope has collapsed, showing the mean velocity vector field over the course of 4 orbits which correspond to those used to produce the mass flux plots in Fig.~\ref{fig:postflow}. The outer circle corresponds to the spherical surface of Fig.~\ref{fig:postflow} at 0.03~\rhill, whilst the inner filled circle illustrates the radius of the planetary core. This slice in the z-${\rm r_{cyl}}$ plane rotates about the core at the same average angular velocity as the gas, which exhibits solid-body rotation within 0.04~\rhill, so that the vector field is seen in the rotating frame of the gas. The bold sections of the outer circle cover the latitudes of $45-70$ degrees at which out flow was seen in the final panel of Fig.~\ref{fig:postflow}.}
\label{fig:convection}
\end{figure}

Fig.~\ref{fig:convection} illustrates the vector field in a slice through the protoplanet, averaged over 4 orbits that correspond to those used to produce the mass flux rendering of Fig.~\ref{fig:postflow}. Within $\sim 0.04$~\rhill \ the protoplanet rotates as a solid body, with a rate of $4.65 \times 10^{-7}~{\rm rad \: s^{-1}}$ (giving a day equivalent to $\approx 160$ Earth days). The plane to which the velocity vectors have been mapped was rotated at this rate, such that the inner region of the atmosphere remained consistently aligned with the plane over the time of averaging. There are four distinct regions of turn over in which gas flows outwards at mid-latitudes before falling inwards again along the disc midplane. These looping flows are subsonic, with mean velocities of less than $3~{\rm m \: s^{-1}}$ and a maximum of less than $8~{\rm m \: s^{-1}}$ in a region where the sound speed is greater than $6~{\rm km \: s^{-1}}$. A time average to a fixed plane, or a spatial average rotating about the z-axis at a single moment in time both reveal a similar structure, suggesting that the general form of these structures is persistent. The bold sections of the 0.03~\rhill \ circle are marked between $45 - 70$~degrees, the latitudes where outflow was seen in the fourth panel of Fig.~\ref{fig:postflow}. As might be expected, the velocity vectors across these bold segments indicate outflow, illustrating that these looping flows are responsible for the features in the rendered plot. The 0.03~\rhill \ region marked by the outer circle contains $\approx 97$ per cent of the total protoplanetary mass, that is the mass measured out to the Hill radius; prior to collapse the same region had contained just under 40 per cent of the mass within the Hill radius. { In this dense environment, possessing a significant radial temperature gradient, the 4 distinct cells revealed in the velocity field might well be indications of convection.}

\begin{figure*}
\centering
\subfigure 
{
    \includegraphics[width=7cm]{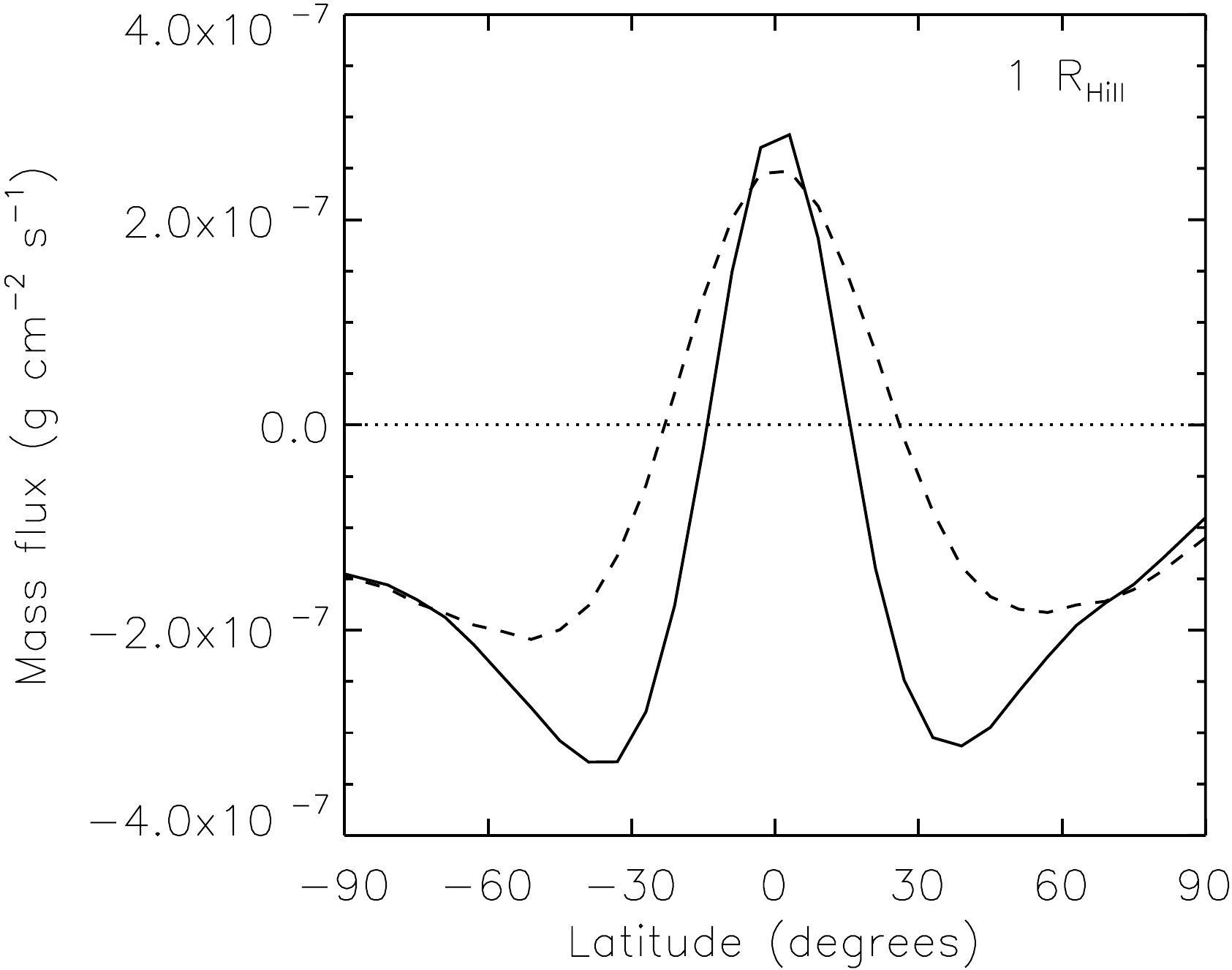}
}
\subfigure 
{
    \includegraphics[width=7cm]{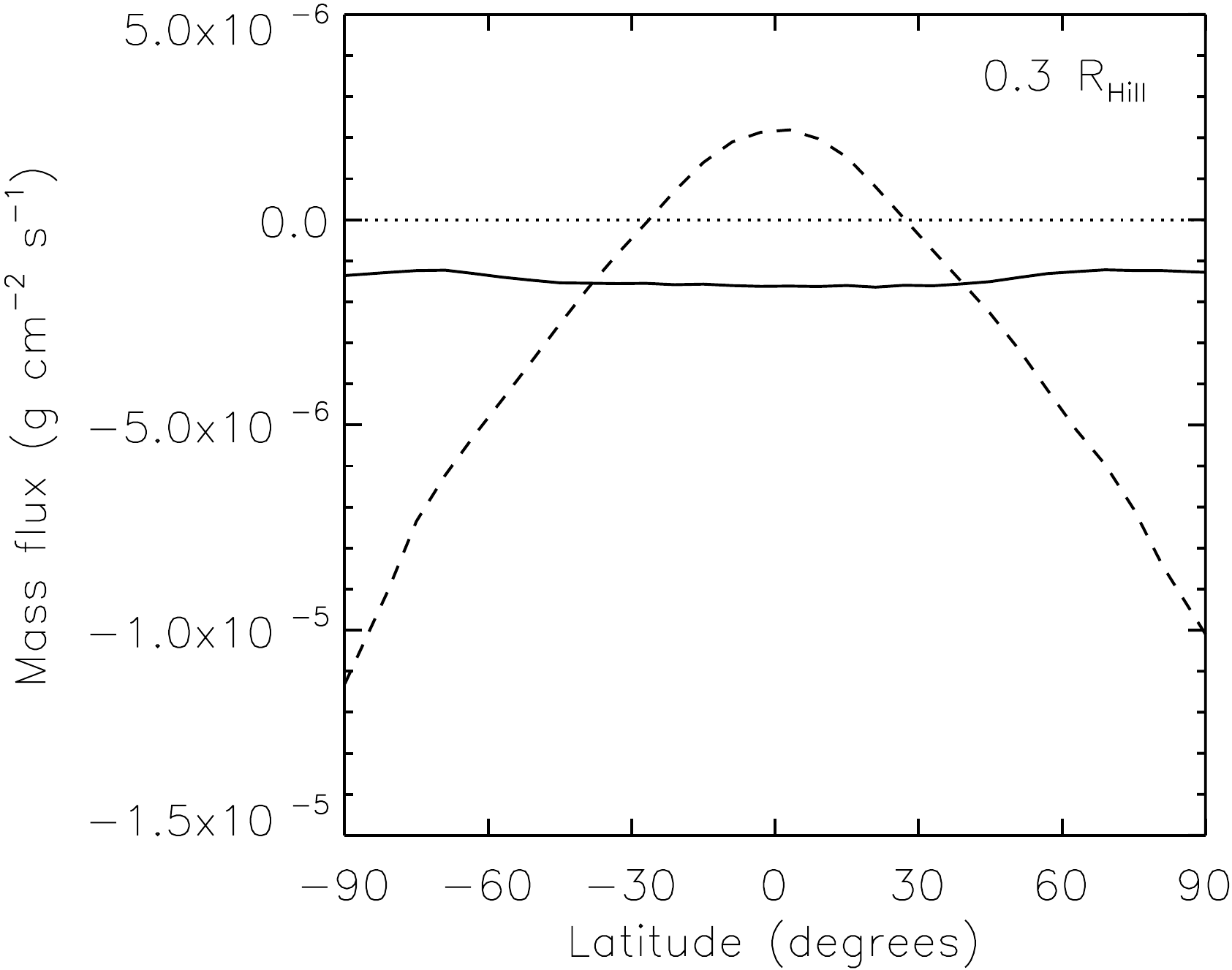}
\vspace{-6pt}
}
\subfigure 
{
    \includegraphics[width=7cm]{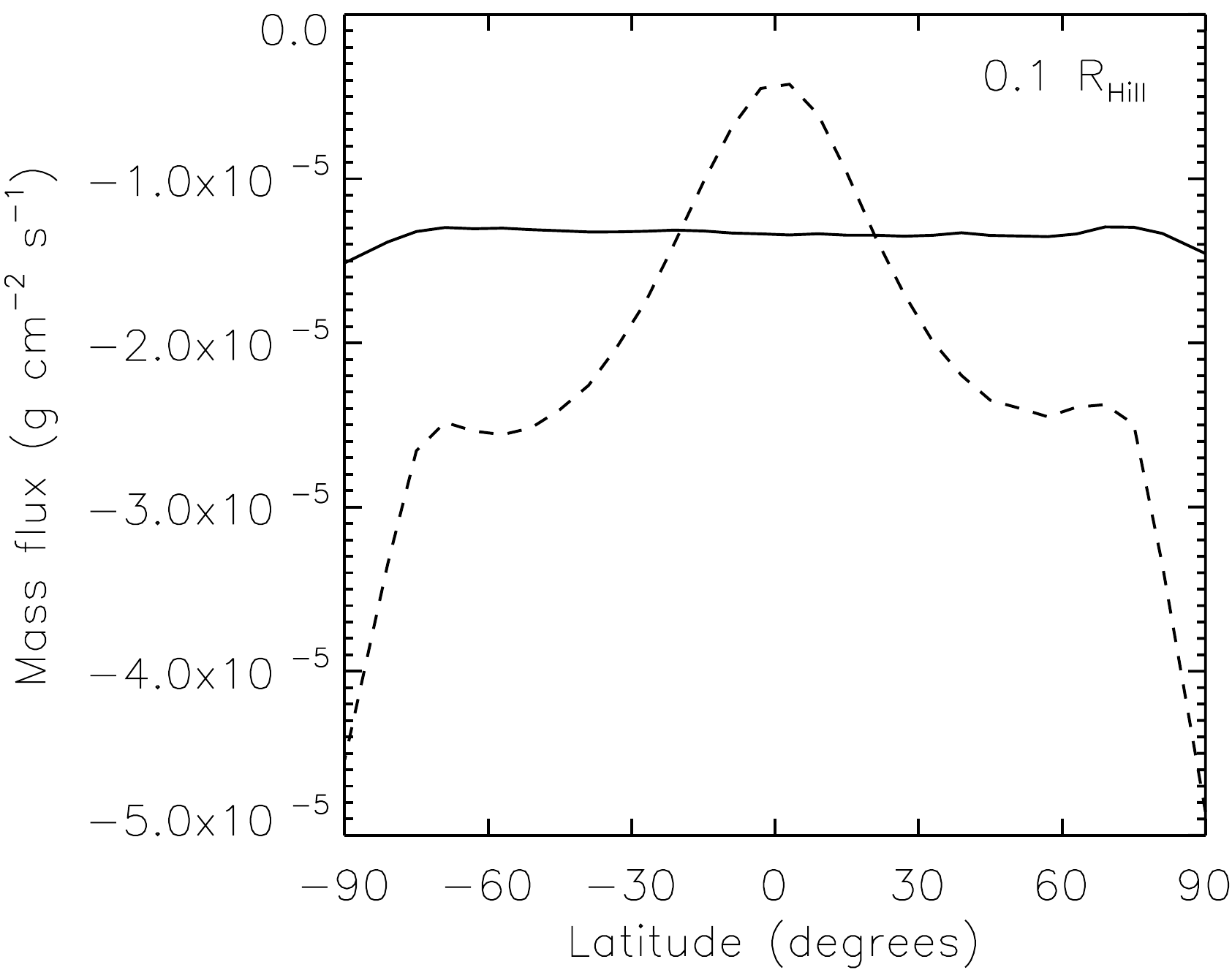}
}
\subfigure 
{
    \includegraphics[width=7cm]{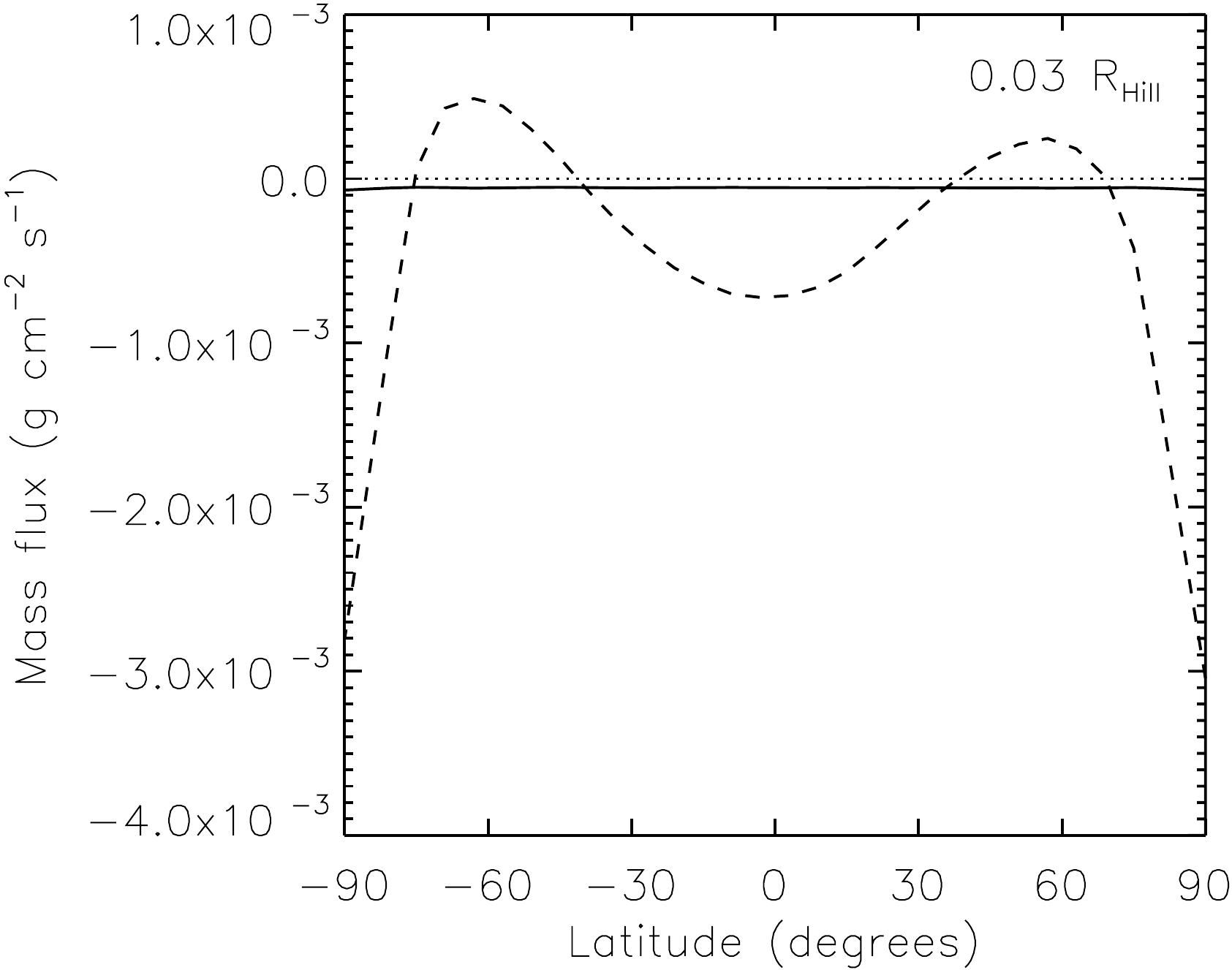}
}

\caption{Longitudinally (azimuthally) integrated mass flux for Model J at the same four radii considered in Figs.~\ref{fig:preflow} \& \ref{fig:postflow}. The solid lines represent the pre-collapse case (Fig.~\ref{fig:preflow}) and the dashed lines the post-collapse case (Fig.~\ref{fig:postflow}). In the former, it is notable that the only instance in which an outward flux can be seen is for the largest radii, where this is easily understood by consideration of the large scale flows (spiral shocks, horseshoe regions, and vertical accretion) surrounding the planet, and persists in a similar form post-collapse. Post-collapse there is more structure visible in the flow at small radii. At 0.3~\rhill \ there is a midplane outflow that corresponds to that visible the corresponding panel of Fig.~\ref{fig:postflow}, whilst at 0.1~\rhill \ all material is flowing inwards, though at the midplane the flux is small relative to other latitudes. Finally, at 0.03~\rhill \ there is evidence of significant turnover in the bound envelope post-collapse, possibly indicative of convection.}
\label{fig:thetaboth}
\end{figure*}

\subsection{Gas accretion}

\cite{TanOhtMac2012} have recently examined gas flow onto and within circumplanetary discs. They found, in agreement with the work of \cite{MacKokInuMat2008}, that gas flowed onto a circumplanetary disc predominantly in the vertical direction. Moreover, their works suggests that material is flowing out along the midplane of a circumplanetary disc. Once the envelope collapses in Model J of this work, and a circumplanetary disc forms, we also find that material is flowing outwards along the midplane beyond a radius of 0.17~\rhill, which is at around half the outer radius of the circumplanetary disc \citep{QuiTri1998, AylBat2009a, MarLub2011}. Fig.~\ref{fig:thetaboth}, which is equivalent to Fig.~6 in \cite{TanOhtMac2012}, illustrates the longitudinally (equivalently, azimuthally) integrated mass flux for the pre (solid line) and post (dashed line) collapse flows shown in Figs.~\ref{fig:preflow} \& \ref{fig:postflow}. Before the envelope collapses, at 0.3~\rhill \ the outflow seen in two places at the midplane is counterbalanced by the associated midplane inflows and the inflow at higher latitudes, yielding a net inflow, as shown by the solid line in the second panel of Fig.~\ref{fig:thetaboth}. However, post collapse the consistent midplane outflow seen for this radius at all longitudes results in a peak of outflowing material in a region $\pm 30$~degrees latitude.

About the planet, at one Hill radius the flow is dominated by the streams of material passing in and out of the region due to the form of their circumstellar orbits (see Fig.~\ref{fig:vfield}). As such, it is unsurprising that the fluxes we obtain, and those of \cite{TanOhtMac2012} scaled to our model, are similar at this radius; note we make the following comparisons using our post-collapse case which better resembles \citeauthor{TanOhtMac2012}'s models. At 1~\rhill \ they obtain a peak outflow along the midplane of $1.8 \times 10^{-7}~{\rm g \: cm^{-2} \: s^{-1}}$, whilst we obtain a value of $2.5 \times 10^{-7}~{\rm g \: cm^{-2} \: s^{-1}}$. The form is also similar, with wings of inflow at higher latitudes of similar magnitude; at the highest latitudes, our normalisation by area leads to a non-zero inflow. However, at smaller radii the mass fluxes we find are considerably larger than at the Hill radius, which differs substantially from \citeauthor{TanOhtMac2012}'s results, where the peak flux at all radii differ by less than an a factor of 4. At 0.3~\rhill \ the peak outward flux has grown to $2.2 \times 10^{-6}~{\rm g \: cm^{-2} \: s^{-1}}$, and at 0.1~\rhill \ the outflow has ceased, but the inflow flux at high latitudes has increased by another factor of around 5. The final panel of Fig.~\ref{fig:thetaboth} reveals the mass flow resulting from the formation of convection cells in the deep atmosphere post collapse.

\section{Summary}

We have performed three-dimensional self-gravitating radiation hydrodynamical models of planet growth that, subsequent to an extended period of growth, result in a dynamical collapse. A series of models were performed using different core masses, core radii, and opacities, that extend the range of accreted mass achieved in \cite{AylBat2009}. We present the first results from a three-dimensional hydrodynamical model of planet growth by core accretion that has been found to produce a hydrodynamic collapse. The result of this collapse is a very centrally-condensed protoplanet, surrounded by a circumplanetary disc, that continues to accrete. The inner reaches of the envelope have undergone significant dissociation of molecular hydrogen, and appear to possess convection-like cells of gas turn over, whilst the inflow of new gas occurs near vertically at high latitudes.

The circumplanetary disc, with radius $\approx \mrhill/3$ and dimensionless scaleheight of $0.4 - 0.5$, exhibits a reversal in the direction of mass flow along the midplane at around 50 per cent of its radius; that is to say there is inflow only within the inner 0.17~\rhill. The degree of central condensation in the post-collapse state leads the model to show good agreement with previous calculations considering mass flow that have presumed a pre-existing high mass core of relatively small size \citep{TanOhtMac2012}. Conversely, before the collapse, the extended protoplanetary envelope exhibits a more spherically symmetric inflow of material.

{ To achieve rapid growth in these models we have adopted very favourable conditions, particularly a low opacity of just 0.1 per cent of the interstellar grain opacity. It may be that the hydrodynamic collapse found in this work is a result of the very rapid growth of the envelope, promoted by these disc conditions, though this cannot be said definitively without performing further models in less favourable discs.}

\section*{Acknowledgments}

We would like to thank the anonymous referee, whose comments helped us to clarify our results. We would also like to thank Peter Bodenheimer for his comments on the manuscript. The calculations reported here were performed using the University of Exeter Supercomputer. Several figures were created using SPLASH \citep{Pri2007}, a visualisation tool for SPH that is publicly available at http://users.monash.edu.au/~dprice/splash/. MRB is grateful for the support of a Philip Leverhulme Prize and a EURYI Award which also supported BAA. This work, conducted as part of the award ``The formation of stars and planets: Radiation hydrodynamical and magnetohydrodynamical simulations"  made under the European Heads of Research Councils and European Science Foundation EURYI (European Young Investigator) Awards scheme, was supported by funds from the Participating Organisations of EURYI and the EC Sixth Framework Programme.

\bibliography{bibliography}

\begin{thebibliography}{48}
\expandafter\ifx\csname natexlab\endcsname\relax\def\natexlab#1{#1}\fi

\bibitem[{{Alexander}(1975)}]{Ale1975}
{Alexander} D.~R., 1975, \apjs, 29, 363

\bibitem[{{Alibert} {et~al.}(2005){Alibert}, {Mordasini}, {Benz}, \&
  {Winisdoerffer}}]{AliMorBenWin2005}
{Alibert} Y., {Mordasini} C., {Benz} W., {Winisdoerffer} C., 2005, \aap, 434,
  343

\bibitem[{{Ayliffe} \& {Bate}(2009)}]{AylBat2009b}
{Ayliffe} B.~A., {Bate} M.~R., 2009, in American Institute of Physics
  Conference Series, Vol. 1158, American Institute of Physics Conference
  Series, {Usuda} T., {Tamura} M., {Ishii} M., eds., pp. 219--221

\bibitem[{{Ayliffe} \& {Bate}(2009a)}]{AylBat2009}
{Ayliffe} B.~A., {Bate} M.~R., 2009a, \mnras, 393, 49

\bibitem[{{Ayliffe} \& {Bate}(2009b)}]{AylBat2009a}
{Ayliffe} B.~A., {Bate} M.~R., 2009b, \mnras, 397, 657

\bibitem[{{Bate}(1995)}]{Bat1995}
{Bate} M., 1995, PhD thesis, PhD thesis, Univ.~Cambridge, (1995)

\bibitem[{{Bate} {et~al.}(2003){Bate}, {Lubow}, {Ogilvie}, \&
  {Miller}}]{BatLubOgiMil2003}
{Bate} M.~R., {Lubow} S.~H., {Ogilvie} G.~I., {Miller} K.~A., 2003, \mnras,
  341, 213

\bibitem[{{Benz}(1990)}]{Ben1990}
{Benz} W., 1990, in Numerical Modelling of Nonlinear Stellar Pulsations
  Problems and Prospects, {J.~R.~Buchler}, ed., Kluwer Academic Publishers,
  Dordrecht, The Netherlands, pp. 269--+

\bibitem[{{Benz} {et~al.}(1990){Benz}, {Cameron}, {Press}, \&
  {Bowers}}]{BenCamPreBow1990}
{Benz} W., {Cameron} A.~G.~W., {Press} W.~H., {Bowers} R.~L., 1990, \apj, 348,
  647

\bibitem[{{Bodenheimer} \& {Pollack}(1986)}]{BodPol1986}
{Bodenheimer} P., {Pollack} J.~B., 1986, Icarus, 67, 391

\bibitem[{{Boley} {et~al.}(2007){Boley}, {Hartquist}, {Durisen}, \&
  {Michael}}]{BolHarDurMic2007}
{Boley} A.~C., {Hartquist} T.~W., {Durisen} R.~H., {Michael} S., 2007, \apjl,
  656, L89

\bibitem[{{Bryden} {et~al.}(1999){Bryden}, {Chen}, {Lin}, {Nelson}, \&
  {Papaloizou}}]{BryCheLinNel1999}
{Bryden} G., {Chen} X., {Lin} D.~N.~C., {Nelson} R.~P., {Papaloizou} J.~C.~B.,
  1999, \apj, 514, 344

\bibitem[{{D'Angelo} {et~al.}(2003{\natexlab{a}}){D'Angelo}, {Henning}, \&
  {Kley}}]{DAnHenKle2003}
{D'Angelo} G., {Henning} T., {Kley} W., 2003{\natexlab{a}}, \apj, 599, 548

\bibitem[{{D'Angelo} {et~al.}(2003{\natexlab{b}}){D'Angelo}, {Kley}, \&
  {Henning}}]{DAnKleHen2003}
{D'Angelo} G., {Kley} W., {Henning} T., 2003{\natexlab{b}}, \apj, 586, 540

\bibitem[{{Hubbard} \& {Macfarlane}(1980)}]{HubMac1980}
{Hubbard} W.~B., {Macfarlane} J.~J., 1980, \jgr, 85, 225

\bibitem[{{Hubickyj} {et~al.}(2005){Hubickyj}, {Bodenheimer}, \&
  {Lissauer}}]{HubBodLis2005}
{Hubickyj} O., {Bodenheimer} P., {Lissauer} J.~J., 2005, Icarus, 179, 415

\bibitem[{{Ikoma} {et~al.}(2000){Ikoma}, {Nakazawa}, \&
  {Emori}}]{IkoNakEmo2000}
{Ikoma} M., {Nakazawa} K., {Emori} H., 2000, \apj, 537, 1013

\bibitem[{{Klahr} \& {Kley}(2006)}]{KlaKle2006}
{Klahr} H., {Kley} W., 2006, \aap, 445, 747

\bibitem[{{Laibe} {et~al.}(2012){Laibe}, {Gonzalez}, \&
  {Maddison}}]{LaiGonMad2012}
{Laibe} G., {Gonzalez} J.-F., {Maddison} S.~T., 2012, \aap, 537, A61

\bibitem[{{Lissauer} {et~al.}(2009){Lissauer}, {Hubickyj}, {D'Angelo}, \&
  {Bodenheimer}}]{LisHubDAnBod2009}
{Lissauer} J.~J., {Hubickyj} O., {D'Angelo} G., {Bodenheimer} P., 2009,
  \icarus, 199, 338

\bibitem[{{Lubow} {et~al.}(1999){Lubow}, {Seibert}, \&
  {Artymowicz}}]{LubSeiArt1999}
{Lubow} S.~H., {Seibert} M., {Artymowicz} P., 1999, \apj, 526, 1001

\bibitem[{{Machida} {et~al.}(2008){Machida}, {Kokubo}, {Inutsuka}, \&
  {Matsumoto}}]{MacKokInuMat2008}
{Machida} M.~N., {Kokubo} E., {Inutsuka} S., {Matsumoto} T., 2008, \apj, 685,
  1220

\bibitem[{{Martin} \& {Lubow}(2011)}]{MarLub2011}
{Martin} R.~G., {Lubow} S.~H., 2011, \mnras, 413, 1447

\bibitem[{{Mizuno}(1980)}]{Miz1980}
{Mizuno} H., 1980, Progress of Theoretical Physics, 64, 544

\bibitem[{{Mizuno} {et~al.}(1978){Mizuno}, {Nakazawa}, \&
  {Hayashi}}]{MizNakHay1978}
{Mizuno} H., {Nakazawa} K., {Hayashi} C., 1978, Progress of Theoretical
  Physics, 60, 699

\bibitem[{{Monaghan}(1992)}]{Mon1992}
{Monaghan} J.~J., 1992, \araa, 30, 543

\bibitem[{{Monaghan}(2002)}]{Mon2002}
{Monaghan} J.~J., 2002, \mnras, 335, 843

\bibitem[{{Monaghan} \& {Gingold}(1983)}]{MonGin1983}
{Monaghan} J.~J., {Gingold} R.~A., 1983, Journal of Computational Physics, 52,
  374

\bibitem[{{Movshovitz} {et~al.}(2010){Movshovitz}, {Bodenheimer}, {Podolak}, \&
  {Lissauer}}]{MovBodPodLis2010}
{Movshovitz} N., {Bodenheimer} P., {Podolak} M., {Lissauer} J.~J., 2010,
  \icarus, 209, 616

\bibitem[{{Paardekooper} \& {Mellema}(2008)}]{PaaMel2008}
{Paardekooper} S., {Mellema} G., 2008, \aap, 478, 245

\bibitem[{{Papaloizou} \& {Nelson}(2005)}]{PapNel2005}
{Papaloizou} J.~C.~B., {Nelson} R.~P., 2005, \aap, 433, 247

\bibitem[{{Perri} \& {Cameron}(1974)}]{PerCam1974}
{Perri} F., {Cameron} A.~G.~W., 1974, Icarus, 22, 416

\bibitem[{{Podolak}(2003)}]{Pod2003}
{Podolak} M., 2003, Icarus, 165, 428

\bibitem[{{Pollack} {et~al.}(1996){Pollack}, {Hubickyj}, {Bodenheimer},
  {Lissauer}, {Podolak}, \& {Greenzweig}}]{PolHubBodLis1996}
{Pollack} J.~B., {Hubickyj} O., {Bodenheimer} P., {Lissauer} J.~J., {Podolak}
  M., {Greenzweig} Y., 1996, Icarus, 124, 62

\bibitem[{{Pollack} {et~al.}(1985){Pollack}, {McKay}, \&
  {Christofferson}}]{PolMcKChr1985}
{Pollack} J.~B., {McKay} C.~P., {Christofferson} B.~M., 1985, Icarus, 64, 471

\bibitem[{{Price}(2007)}]{Pri2007}
{Price} D.~J., 2007, Publications of the Astronomical Society of Australia, 24,
  159

\bibitem[{{Price} \& {Monaghan}(2007)}]{PriMon2007}
{Price} D.~J., {Monaghan} J.~J., 2007, \mnras, 374, 1347

\bibitem[{{Quillen} \& {Trilling}(1998)}]{QuiTri1998}
{Quillen} A.~C., {Trilling} D.~E., 1998, \apj, 508, 707

\bibitem[{{Sasaki}(1989)}]{Sas1989}
{Sasaki} S., 1989, \aap, 215, 177

\bibitem[{{Seager} {et~al.}(2007){Seager}, {Kuchner}, {Hier-Majumder}, \&
  {Militzer}}]{SeaKucHieMil2007}
{Seager} S., {Kuchner} M., {Hier-Majumder} C.~A., {Militzer} B., 2007, \apj,
  669, 1279

\bibitem[{{Slattery}(1977)}]{Sla1977}
{Slattery} W.~L., 1977, Icarus, 32, 58

\bibitem[{{Springel} \& {Hernquist}(2002)}]{SprHer2002}
{Springel} V., {Hernquist} L., 2002, \mnras, 333, 649

\bibitem[{{Tajima} \& {Nakagawa}(1997)}]{TajNak1997}
{Tajima} N., {Nakagawa} Y., 1997, Icarus, 126, 282

\bibitem[{{Tanigawa} {et~al.}(2012){Tanigawa}, {Ohtsuki}, \&
  {Machida}}]{TanOhtMac2012}
{Tanigawa} T., {Ohtsuki} K., {Machida} M.~N., 2012, \apj, 747, 47

\bibitem[{{Whitehouse} \& {Bate}(2006)}]{WhiBat2006}
{Whitehouse} S.~C., {Bate} M.~R., 2006, \mnras, 367, 32

\bibitem[{{Whitehouse} {et~al.}(2005){Whitehouse}, {Bate}, \&
  {Monaghan}}]{WhiBatMon2005}
{Whitehouse} S.~C., {Bate} M.~R., {Monaghan} J.~J., 2005, \mnras, 364, 1367

\bibitem[{{Wuchterl}(1990)}]{Wuc1990}
{Wuchterl} G., 1990, \aap, 238, 83

\bibitem[{{Wuchterl}(1991)}]{Wuc1991}
{Wuchterl} G., 1991, Icarus, 91, 53

\end{thebibliography}

\end{document}